\documentclass[11pt]{article}

\usepackage{epsfig,amsmath,latexsym,amssymb}
\usepackage{graphicx}
\usepackage{lscape}
\usepackage{picture, eso-pic, tikz} 
\usepackage{dsfont}
\usepackage{listings}
\usepackage{changes}
\usepackage{hyperref}
\usepackage{bbding}

\renewcommand{\baselinestretch}{1.1}
\def\R{{\mathbb R}}  
\def\N{{\mathbb N}}  
\def\p{{\mathbb P}}  
\def\E{{\mathbb E}}  %
\def\Beweis{\normalsize}

\newcommand{\Remm}[1]{}
\newtheorem{theo}{Theorem}[section]

\newtheorem{defi}[theo]{Definition}
\newtheorem{model ass}[theo]{Model Assumptions}

\newtheorem{rem}[theo]{Remark}
\newtheorem{rems}[theo]{Remarks}

\def\EndProof{\hfill {\scriptsize $\Box$}}

\numberwithin{equation}{section}

\definecolor{MyGray}{rgb}{0.92,0.92,0.92}


\definecolor{British racing}{rgb}{0.0, 0.5, 0.0}
\def\by{\boldsymbol{y}}
\def\bx{\boldsymbol{x}}

\def\bz{\boldsymbol{z}}

\def\bY{\boldsymbol{Y}}
\def\bX{\boldsymbol{X}}

\def\b0{\boldsymbol{0}}

\def\balpha{\boldsymbol{\alpha}}
\def\boldeta{\boldsymbol{\eta}}
\def\bmu{\boldsymbol{\mu}}

\def\bepsilon{\boldsymbol{\epsilon}}

\def\b0{\boldsymbol{0}}

\usepackage{todonotes}
\newcommand{\Comments}{1}
\newcommand{\mynote}[2]{\ifnum\Comments=1\textcolor{#1}{#2}\fi}
\newcommand{\mytodo}[2]{\ifnum\Comments=1%
  \todo[linecolor=#1!80!black,backgroundcolor=#1,bordercolor=#1!80!black]{#2}\fi}

\ifnum\Comments=1               
\fi

\ifnum\Comments=1               
\fi

\ifnum\Comments=1               
\fi

\begin{document}
\author{Mario V.~W\"uthrich\footnote{RiskLab, Department of Mathematics, ETH Zurich,
mario.wuethrich@math.ethz.ch} 
\and Johanna Ziegel\footnote{Institute of Mathematical Statistics and Actuarial Science, University of Bern, johanna.ziegel@stat.unibe.ch}}

\date{Version of \today}
\title{Isotonic Recalibration under a Low Signal-to-Noise Ratio}
\maketitle

\begin{abstract}
\noindent  
Insurance pricing systems should fulfill the auto-calibration property to ensure that there is no systematic cross-financing between different price cohorts. Often, regression models are not auto-calibrated. We propose to apply isotonic recalibration to a given regression model to ensure auto-calibration. Our main result proves that under a low signal-to-noise ratio, this isotonic recalibration step leads to explainable pricing systems because the resulting isotonically recalibrated regression functions have a low complexity.

~

\end{abstract}

\section{Introduction}
There are two seemingly unrelated problems in insurance pricing that we are going to tackle in this paper. First, an insurance pricing system should not have any systematic cross-financing between different price cohorts. Systematic cross-financing implicitly means that some parts of the portfolio are under-priced, and this is compensated by other parts of the portfolio that are over-priced. We can prevent systematic cross-financing between price cohorts by ensuring that the pricing system is {\it auto-calibrated}. We propose to apply {\it isotonic recalibration} which turns any regression function into an auto-calibrated pricing system.

The second problem that we tackle is the explainability of complex algorithmic models for insurance pricing. In a first step, one may use any complex regression model to design an insurance pricing system such as, e.g., a deep neural network. Such complex regression models typically lack explainability and rather act as black boxes. For this reason, there are several tools deployed to explain such complex solutions, we mention, for instance, SHAP by Lundberg--Lee \cite{LundbergLee}. Since algorithmic solutions do not generally fulfill the aforementioned auto-calibration property, we propose to apply isotonic recalibration to the algorithmic solution. If the signal-to-noise ratio is low in the data, then the isotonic recalibration step leads to a coarse partition of the covariate space and, as a consequence, it leads to an explainable version of the algorithmic model used in the first place. Thus, explainability is a nice side result of applying isotonic recalibration in low signal-to-noise ratio problems, which is typically the case in insurance pricing settings. 

There are other methods for obtaining auto-calibration through a recalibration step; we mention Lindholm et al.~\cite{Lindholm} and Denuit et al.~\cite{DenuitCharpentierTrufin}. These other methods often require tuning of hyperparameters, e.g., using cross-validation. Isotonic recalibration does not involve any hyperparameters as it solves a constraint regression problem (ensuring monotonicity). As such, isotonic recaliabration is universal because it also does not depend on the specific choice of the loss function within the family of Bregman losses.

\medskip

We formalize our proposal.
Throughout, we assume that all considered random variables have finite means.
Consider a response variable $Y$ that is equipped with covariate information $\bX \in  {\cal X} \subseteq \R^q$.
The general goal is to determine the (true) regression function $\bx \mapsto \E[ Y| \bX = \bx]$ that
describes the conditional mean of $Y$, given $\bX$. Typically, this true regression function is
unknown, and it needs to be determined from i.i.d.~data $(y_i, \bx_i)_{i=1}^n$, that is, a sample from $(Y,\bX)$. For this purpose, we try to select a regression function $\bx \mapsto \mu(\bx)$ from a 
(pre-chosen) function class on ${\cal X}$ that approximates
the conditional mean $\E[ Y| \bX = \cdot]$ as well as possible. Often, it is not possible to capture all features of the regression function from data. In financial applications, a minimal important requirement
for a well-selected regression function $\mu(\cdot)$ is that it fulfills the auto-calibration
property.
\begin{defi} The regression function $\mu$ is auto-calibrated for $(Y,\bX)$ if
  \begin{equation*}
    \mu(\bX) = \E \left[ \left. Y \right| \mu(\bX) \right],
    \qquad \text{$\p$-a.s.}
  \end{equation*}
\end{defi}
Auto-calibration is an important property in actuarial and financial applications because it implies that, on average, the (price) cohorts
$\mu(\bX)$ are self-financing for the corresponding claims $Y$, i.e., there is no systematic cross-financing
within the portfolio, if the structure of this portfolio is described by the covariates $\bX \sim \p$ and the price
cohorts $\mu(\bX)$, respectively. In a Bernoulli context, an early version of auto-calibration (called well-calibrated)
has been introduced by Schervish \cite{Schervish} to the community in statistics, and recently, it has been considered in detail by Gneiting--Resin \cite{GneitingResin2021}. In an actuarial and financial context,
the importance of auto-calibration has been emphasized in Kr\"uger--Ziegel \cite{Ziegel},
Denuit et al.~\cite{DenuitCharpentierTrufin}, W\"uthrich \cite{WGini} and  Lindholm et al.~\cite{Lindholm}.

Many regression models do not satisfy the auto-calibration property. However, there is a simple and powerful method, which we call \emph{isotonic recalibration}, to obtain an (in-sample) auto-calibrated regression function starting from any candidate function $\pi:\mathcal{X} \to \R$. We apply isotonic recalibration to the pseudo-sample $(y_i,\pi(\bx_i))_{i=1}^n$ to obtain an isotonic regression function $\widehat{\mu}$. Then,
\begin{equation}\label{eq:auto-in-sample}
\widehat{\mu}(\bX') = \E \left[Y'| \widehat{\mu}(\bX')\right], \quad \text{$\p_n$-a.s.,}
\end{equation}
where $(Y',\bX')$ is distributed according to the empirical distribution $\p_n$ of $(y_i,\bx_i)_{i=1}^n$; see Section \ref{sec:isoreg} for details. Isotonic regression determines an adaptive partition of the covariate space ${\cal X}$, and $\widehat{\mu}$ is determined by averaging $y$-values over the partition elements. Clearly, other binning approaches can also be used on the pseudo-sample $(y_i,\pi(\bx_i))_{i=1}^n$ to enforce \eqref{eq:auto-in-sample}, but we argue that isotonic regression is preferable since it avoids subjective choices of tuning parameters and leads to sensible regression functions under reasonable and verifiable assumptions. The only assumption for isotonic recalibration to be informative is that the function $\pi$ gets the rankings of the conditional means right, that is, whenever $\E \left[Y| \bX=\bx_i\right]\le \E \left[Y| \bX=\bx_j\right]$, we would like to have $\pi(\bx_i) \le \pi(\bx_j)$.

Using isotonic regression for recalibration is not new in the literature. In the case of binary outcomes, it as already been proposed by Zadrozny--Elkan~\cite{ZadroznyElkan2002}, Menon et al.~\cite{Menon} and recently by Tasche \cite[Section 5.3]{Tasche2}. The monotone single index models of Balabdaoui et al.~\cite{BalabdaouiDurotETAL2019} follow the same strategy as described above but the focus of their work is different from ours. They specifically consider a linear regression model for the candidate function $\pi$, which is called the index. In the case of distributional regression, that is, when interest is in determining the whole conditional distribution of $Y$ given covariate information $\bX$, Henzi et al.~\cite{HenziKlegerETAL2021a} have suggested to first estimate an index function $\pi$ that determines the ordering of the conditional distributions w.r.t.~first order stochastic dominance and then estimate conditional distributions using isotonic distributional regression; see Henzi et al.~\cite{HenziZiegelETAL2021}.

As a new contribution, we show that the size of the partition of the isotonic recalibration may give insight concerning the information content of the recalibrated regression function $\widehat{\mu}$. Furthermore, the partition of the isotonic recalibration allows to explain connections between covariates and outcomes, in particular, when the signal-to-noise ratio is small which typically is the case for insurance claims data. 

In order to come up with a candidate function $\pi:{\cal X} \to \R$, one may consider any regression model such as, e.g.,
a generalized linear model, a regression tree, a tree boosting regression model or a deep neural network regression model. The aim is that $\pi(\cdot)$ provides us with the correct rankings of the conditional means $\E[Y|\bX = \bx_i]$, $i=1, \ldots, n$. 
The details are discussed in Section \ref{sec:isouse}. 

\bigskip
{\bf Organization.} In Section \ref{sec: Isotonic regression}, we formally introduce isotonic regression which is a constraint optimization problem. This constraint optimization problem is usually solved with the pool adjacent violators (PAV) algorithm, which is described in Appendix \ref{appendix PAVA}. Our main result is stated in Section \ref{Monotonicity of the expected complexity number}. It relates the 
complexity of the isotonic recalibration
solution to the signal-to-noise ratio in the data. Section \ref{sec:isouse} gives practical guidance on the use of isotonic recalibration, and in Section \ref{sec:examples} we exemplify our results on a frequently used insurance data set. In this section we also present graphic tools for interpreting the regression function. In Section \ref{sec: Conclusions}, we conclude.

\section{Isotonic regression}
\label{sec: Isotonic regression}
\subsection{Definition and basic properties}\label{sec:isoreg}
For simplicity, we assume that the candidate function $\pi:{\cal X}\to \R$ does not lead to any ties in the values $\pi(\bx_1),\dots,\pi(\bx_n)$, and that the indices $i=1,\ldots, n$ are chosen such that they are aligned with the ranking, that is, $\pi(\bx_1) < \ldots < \pi(\bx_n)$. Remark \ref{remark ties} explains how to handle ties.
The isotonic regression of $\bz = (y_i,\pi(\bx_i))_{i=1}^n$ with positive case weights $(w_i)_{i=1}^n$ is the solution $\widehat{\bmu}\in \R^n$ to the restricted minimization problem
\begin{equation}\label{eq:iso1}
  \widehat{\bmu} ~=~\underset{\bmu=(\mu_1,\ldots, \mu_n)^\top}{\arg\min}~
  \sum_{i=1}^n w_i \left(y_i - \mu_i\right)^2, \qquad \text{subject to $\mu_1 \le \ldots \le \mu_n$.}
\end{equation}
We can rewrite the side constraints as $A \bmu \ge \b0$ (component-wise), where $A=(a_{i,j})_{i,j}\in \R^{n \times (n-1)}$
is the matrix with the elements $a_{i,j}=\mathds{1}_{i=j-1}-\mathds{1}_{i=j}$.
We define $\by=(y_1, \ldots, y_n)^\top \in \R^n$ and the (diagonal) case weight matrix $W={\rm diag}(w_1,\ldots, w_n)$.
The above optimization problem then reads as 
\begin{equation}\label{isotonic regression}
  \widehat{\bmu} 
  ~=~\widehat{\bmu}(\bz) ~=~\underset{\bmu:\,A\bmu \ge \b0}{\arg\min}~ 
   (\by - \bmu)^\top W (\by - \bmu).
 \end{equation}
 This shows that the isotonic regression is solved by a convex minimization with linear side constraints. It remains
 to verify that the auto-calibration property claimed in \eqref{eq:auto-in-sample} holds. 

\begin{rem}\label{remark ties}\normalfont
If there are ties in the values $\pi(\bx_1),\dots,\pi(\bx_n)$, for example, $\pi(\bx_i) = \pi(\bx_j)$ for some $i \not=j$, we replace $y_i$ and $y_j$ with their weighted average $(w_iy_i+w_jy_j)/(w_i+w_j)$ and assign them weights $(w_i+w_j)/2$. The procedure is analogous for more than two tied values. This corresponds to the second option of dealing with ties in Leeuw et al.~\cite[Section 2.1]{Leeuw}.
\end{rem}

\begin{rem}\label{rem:Bregman}\normalfont
Barlow et al.~\cite[Theorem 1.10]{BarlowEtAl} show that the square loss function in \eqref{eq:iso1} can be replaced by any Bregman loss function, $L_\phi(y,\mu) = \phi(y) - \phi(\mu) + \phi'(\mu)(y-\mu)$, without changing the optimal solution $\widehat{\bmu}$. Here, $\phi$ is a strictly convex function with subgradient $\phi'$. Bregman loss functions are the only consistent loss functions for the mean; see Savage \cite{Savage} and Gneiting \cite[Theorem 7]{Gneiting}. If $y$ and $\mu$ only take positive values, a Bregman loss function of relevance for this paper is the gamma deviance loss, which is equivalent to the QLIKE loss that arises by choosing $\phi(x) = -\log(x)$; see Patton \cite{Patton2011}.
\end{rem}

The solution to the minimization problem \eqref{isotonic regression} can be given explicitly as a min-max formula, that is,
\[
\widehat{\mu}_i ~=~ \min_{\ell = i,\dots,n} \max_{k=1,\dots,\ell}\, \frac{1}{\sum_{j=k}^\ell w_j}\,\sum_{j=k}^\ell w_j y_j.
\]
While the min-max formula is theoretically appealing and useful, the related minimum lower sets (MLS) algorithm of Brunk et al.~\cite{Brunk} is not efficient to compute the solution. The pool adjacent violators (PAV) algorithm, which is due to Ayer et al.~\cite{Ayer}, Miles \cite{Miles} and Kruskal \cite{Kruskal}, allows for fast computation of the isotonic regression and provides us with the desired insights about the solution. In Appendix \ref{appendix PAVA}, we describe the PAV algorithm in detail. 
The solution is obtained by suitably partitioning the index set ${\cal I}=\{1,\ldots, n\}$
into (discrete) intervals
\begin{equation}\label{discrete interval}
  {\cal I}_k ={\cal I}_k(\bz) =\{i_{k-1}+1, \ldots, i_k \} \qquad \text{ for ~$k=1,\ldots, K(\bz)$,}
\end{equation}
with $\bz$-dependent slicing points $0=i_0< i_1 < \ldots < i_K=n$, and with $K(\bz)\in \{1,\ldots, n\} $ denoting the number of discrete intervals
${\cal I}_k$.
The number $K(\bz)$ of intervals and the slicing points $i_k=i_k(\bz)$, $k=1,\ldots, K(\bz)$, for the partition
of ${\cal I}$ depend on the observations $\bz$. On each discrete interval ${\cal I}_k$
we then obtain the isotonic regression parameter
estimate for instance $i\in {\cal I}_k$
\begin{equation}\label{PAVA block estimate}
      \widehat{\mu}_i= \widehat{\mu}_{i_k}=\frac{1}{\sum_{j \in {\cal I}_k}w_j}\,\sum_{j \in {\cal I}_k}w_jy_j,
\end{equation}
see also \eqref{PAVA block estimate A}.
Thus, on each block ${\cal I}_k$ we have a constant estimate $\widehat{\mu}_{i_k}$, and the isotonic property tells us
that these estimates are strictly increasing over the block indices $k=1,\ldots, K(\bz)$, because these
blocks have been chosen to be maximal. We call $K(\bz)$ the {\it complexity number} of the resulting
isotonic regression. 

\begin{figure}[htb!]
\begin{center}
\begin{minipage}[t]{0.4\textwidth}
\begin{center}
\includegraphics[width=\textwidth]{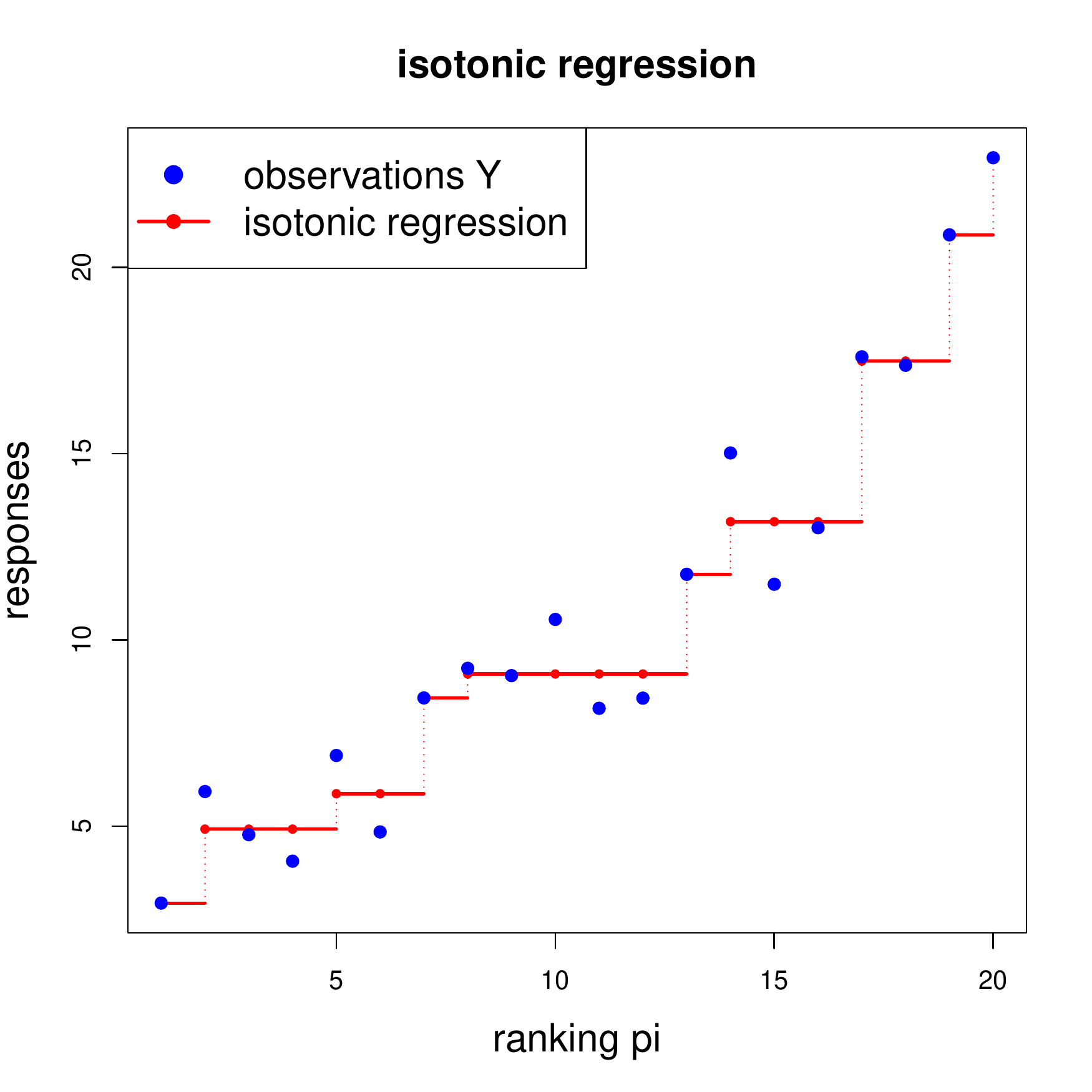}
\end{center}
\end{minipage}
\end{center}
\vspace{-.7cm}
\caption{Example of an isotonic regression with $K(\bz)=10$ blocks.}
\label{fig:simpleiso}
\end{figure}

Figure \ref{fig:simpleiso} gives an example for $n=20$ and rankings $\pi(\bx_i)=i$ for $i=1,\ldots, n$. The resulting (non-parametric)
isotonic regression function $\widehat{\bmu}=\widehat{\bmu}(\bz)$, which is only uniquely determined at the observations $(\pi(\bx_i))_{i=1}^n$, can be interpolated by a step function. In Figure \ref{fig:simpleiso} this results in a step function having $K(\bz)-1=9$ steps,
that is, we have $K(\bz)=10$ blocks, and the estimated regression function $\widehat{\mu}$ takes only $K(\bz)=10$ different values. This motivates to call $K(\bz)$ the complexity number of the resulting step function, see Figure \ref{fig:simpleiso}.

The partition of the indices ${\cal I}$ into the isotonic blocks ${\cal I}_k$ is obtained naturally by requiring
monotonicity. This is different from the regression tree approach considered in Lindholm et al.~\cite{Lindholm}. In fact,
this latter reference does not require monotonicity but aims at minimizing the ``plain'' square loss
using, e.g., cross-validation for determining the optimal number of partitions. In our context, the complexity number $K(\bz)$ is fully determined through requiring monotonicity and, in general, the results will differ.

In insurance applications, the blocks  ${\cal I}_k \subset {\cal I}$ provide us with the (empirical) price cohorts
$\widehat{\mu}_i=\widehat{\mu}_{i_k}$, for $i \in {\cal I}_k$, and \eqref{PAVA block estimate} leads to
the (in-sample) auto-calibration property for $Y$
\begin{equation}\label{auto-calibration step 2}
  \E\left[\left. Y' \right| \widehat{\mu}(\bX')=\widehat{\mu}_{i_k}\right]~=~\frac{1}{\sum_{i \in {\cal I}_k}w_i}\,\sum_{i \in {\cal I}_k}w_iy_i
  ~=~\widehat{\mu}_{i_k},
  \end{equation}
  where $(Y',\bX')$ is distributed according to the weighted empirical distribution of $(y_i,\bx_i)_{i=1}^n$ with weights $(w_i)_{i=1}^n$. 
Moreover, summing over the entire portfolio we have the (global) balance property
\begin{equation}\label{global balance}
    \sum_{i=1}^n w_i \widehat{\mu}_i = \sum_{k=1}^{K(\bz)}\sum_{i \in {\cal I}_k} w_i \widehat{\mu}_i
    = \sum_{k=1}^{K(\bz)} \widehat{\mu}_{i_k} \sum_{i \in {\cal I}_k} w_i
    = \sum_{k=1}^{K(\bz)}  \sum_{i \in {\cal I}_k} w_i y_i =\sum_{i=1}^n w_i y_i,
\end{equation}  
that is, in average the overall (price) level is correctly specified if we price the insurance policies with covariates $\bx_i$
by $w_i \widehat{\mu}_i$, where the weights $w_i>0$ now receive the interpretation of exposures.

\subsection{Monotonicity of the expected complexity number}
\label{Monotonicity of the expected complexity number}

In this section, we prove that the expected complexity number $\E[K(\bz)]$
is an increasing function of the signal-to-noise ratio. For this,
we assume a location-scale model for the responses $Y_i$, that is, we assume that
\begin{equation}\label{signal-to-noise ratio}
  Y_i = \mu_i + \sigma \epsilon_i, \quad i = 1,\dots,n,
\end{equation}
with noise terms $\epsilon_i$, location parameters $\mu_i\in \R$ with $\mu_1 \le \ldots \le \mu_n$, and scale parameter $\sigma >0$. Here, $\mu_i$ takes the role of $\pi(\bx_i)$ in the previous section. The parameters $\mu_1,\dots,\mu_n$ are unknown but it is known that they are labeled in increasing order.
The signal-to-noise ratio is then described by the scale parameter $\sigma$, i.e., we have a low signal-to-noise ratio
for high $\sigma$ and vice-versa. The explicit location-scale structure 
\eqref{signal-to-noise ratio} allows us to analyze
\begin{equation}\label{signal-to-noise ratio 2}
  \by~=~\bY_\sigma(\omega) ~=~ \bmu + \sigma \bepsilon(\omega)
  ~=~(\mu_1,\ldots, \mu_n)^\top + \sigma (\epsilon_1, \ldots, \epsilon_n)^\top(\omega),
\end{equation}   
point-wise in the sample points $\omega \in \Omega$ of the probability space $(\Omega, {\cal F}, \p)$
as a function of $\sigma>0$; this is similar to the re-parametrization trick
of Kingma--Welling \cite{KingmaWelling} that is frequently used to explore
variational auto-encoders.

In this section, we write $K(\by) = K(\bz)$, because
the ranking of the outcomes $\by$ is clear from the context (labeling), and we do not go via a ranking function $\pi(\cdot)$.

\begin{theo}\label{signal-to-noise theorem}
  Assume that the responses $Y_i$, $i=1,\dots,n$, follow the location-scale model \eqref{signal-to-noise ratio}
  with (unknown) ordered location
  parameters $\mu_1 < \ldots < \mu_n$, and scale parameter $\sigma>0$. Then, the expected complexity number $\E[K(\bY)]$ of the isotonic regression of $\bY$ is a decreasing function in $\sigma>0$. If the distribution of the noise vector $\bepsilon=(\epsilon_1,\dots,\epsilon_n)^{\top}$ has full support on $\R^n$, then $\E[K(\bY)]$  is strictly decreasing in $\sigma$.
\end{theo}  

Theorem \ref{signal-to-noise theorem} proves that, under a specific but highly relevant model, the complexity number $K(\bY)$ of the isotonic regression is decreasing on average with a decreasing
signal-to-noise ratio. Implicitly, this means that more noisy data, which has a lower
information ratio, leads to a less granular regression function. Consequently, if the partition of the isotonic regression is used to obtain a partition of the covariate space ${\cal X}$ via the candiate function $\pi$, this partition will be less granular, the more noise of $\bY$ cannot be explained by $\pi(\bX)$, see also Section \ref{sec:diagnostics} for a further discussion. 

To the best of our knowledge, our result is a new contribution to the literature on isotonic regression. While we focus on the finite sample case, a related result is the analysis of the complexity number of the isotonic regression function as function of the sample size $n$, see Dimitriadis et al.~\cite[Lemma 3.2]{Dimitriadis}. 

We are assuming strictly ordered location parameters in the formulation of Theorem \ref{signal-to-noise theorem}. This assumption simplifies the proof in the case where we show that the expected complexity number $K(\bY)$ is strictly decreasing in $\sigma$. With some additional notation, the theorem could be generalized to allow for ties between some (but not all) $\mu_i$. 

\begin{figure}[htb!]
\begin{center}
\begin{minipage}[t]{0.4\textwidth}
\begin{center}
\includegraphics[width=\textwidth]{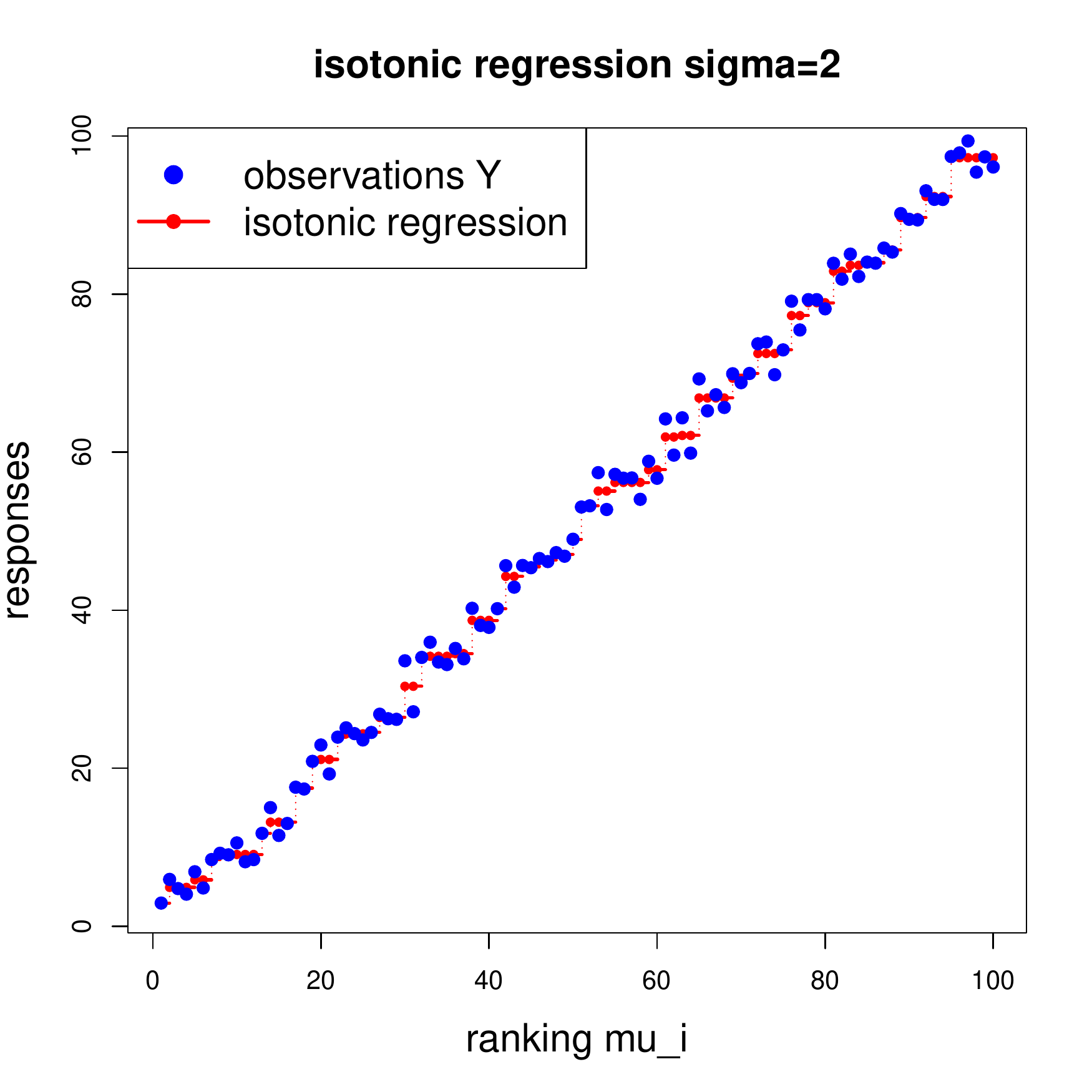}
\end{center}
\end{minipage}
\begin{minipage}[t]{0.4\textwidth}
\begin{center}
\includegraphics[width=\textwidth]{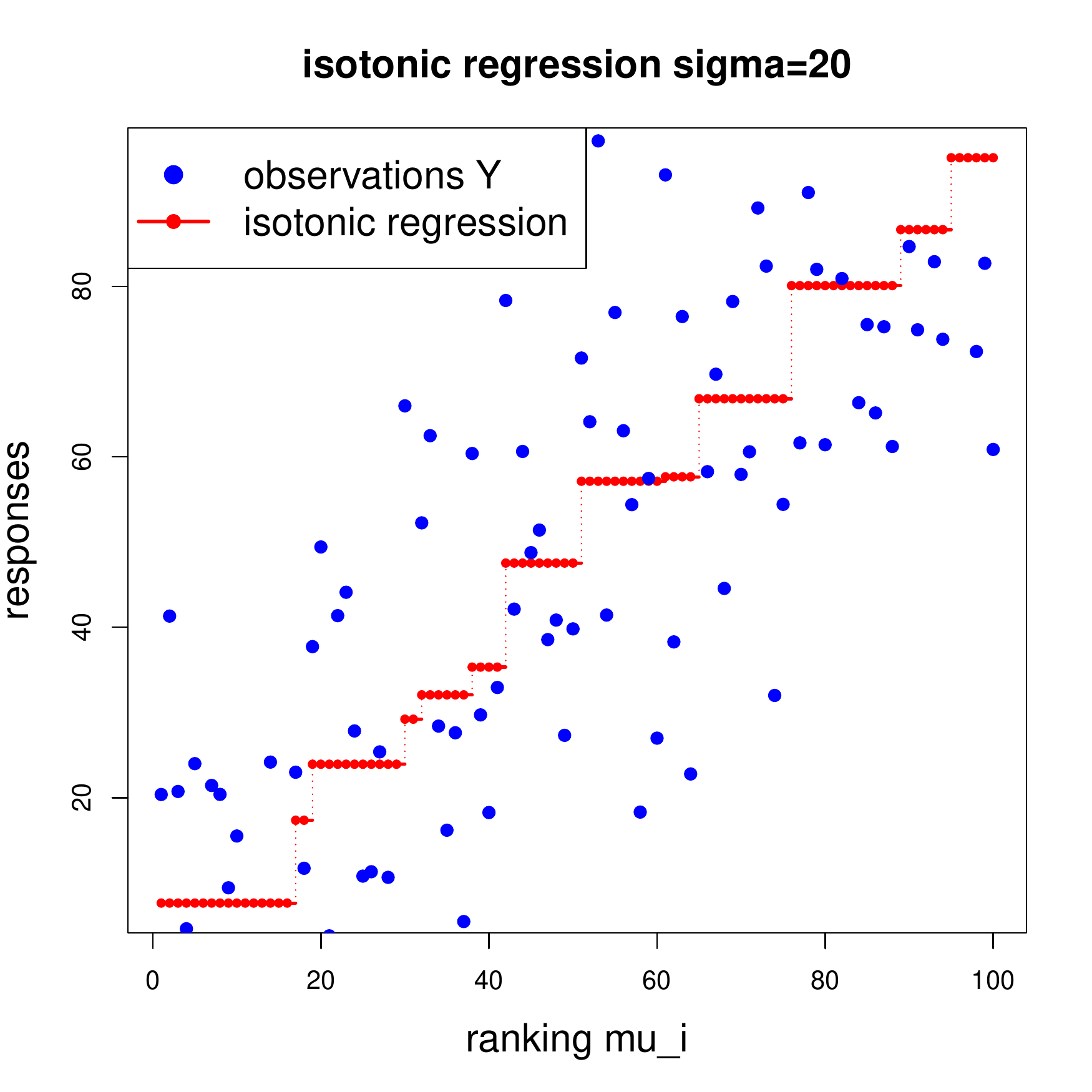}
\end{center}
\end{minipage}
\end{center}
\vspace{-.7cm}
\caption{Example of an isotonic regression of location-scale type with varying signal-to-noise ratio
  for the identical sample point $\omega \in \Omega$: (lhs)
  $\sigma=2$ with $K(\by)=46$ and (rhs) $\sigma=20$ with $K(\by)=13$.}
\label{fig: Example 2}
\end{figure}

Figure \ref{fig: Example 2} gives an example of a location-scale model \eqref{signal-to-noise ratio} with i.i.d.~standard
Gaussian noise and scale parameters $\sigma=2$ (lhs) and $\sigma=20$ (rhs), and both figures
consider the same sample point $\omega \in \Omega$ in the noise term $\bepsilon(\omega)$, see \eqref{signal-to-noise ratio 2}.
On the right-hand side of Figure \ref{fig: Example 2}, we have complexity number $K(\by)=13$, and on
the left-hand side $K(\by)=46$; the chosen sample size is $n=100$.

\section{Isotonic recalibration for prediction and interpretation}\label{sec:isouse}

\subsection{Prediction and estimation}

In order to determine an auto-calibrated model for the true regression function $\bx \mapsto \E[Y|\bX = \bx]$ from i.i.d.~data $(y_i,\bx_i)_{i=1}^n$, we are suggesting a two-step estimation procedure. First, we choose a regression model and use the data $(y_i,\bx_i)_{i=1}^n$ to obtain an estimate $\widehat{\pi}$ of a candidate function $\pi$ that should satisfy
\begin{equation}\label{eq:iso-assumption}
\pi(\bx) \le \pi(\bx') \quad \Longleftrightarrow\quad \E[Y| \bX = \bx] \le \E[Y| \bX = \bx'], 
\end{equation}
for all $\bx,\bx' \in \mathcal{X}$. For example, in the case study in Section \ref{sec:examples}, a deep neural network model is chosen for $\pi$. For sensible results, it is important that the estimation method for $\widehat{\pi}$ does not overfit to the data. 

In the second step, we apply isotonic regression to the pseudo-sample  $(y_i,\widehat{\pi}(\bx_i))_{i=1}^n$ 
to obtain an in-sample auto-calibrated regression function $\widehat{\mu}$ defined on $\{\widehat{\pi}(\bx_i):i=1,\dots,n\}$. We call this second step \emph{isotonic recalibration}. In order to obtain a prediction for a new covariate value $\bx \in \mathcal{X}$, we compute $\widehat{\pi}(\bx)$, find $i$ such that $\widehat{\pi}(\bx_{i}) < \widehat{\pi}(\bx) \le \widehat{\pi}(\bx_{i+1})$, and interpolate by setting $\widehat{\mu}(\bx) = (\widehat{\mu}(\bx_{i})+ \widehat{\mu}(\bx_{i+1}))/2$. This interpolation may be advantageous for prediction. For interpretation and analysis, however, we prefer a step function interpolation as this leads to a partition of the covariate space, see Section \ref{sec:diagnostics}, below, and Figure \ref{fig: Example 2}. 

This two-step estimation approach can be interpreted as a generalization of the monotone single index models considered by Balabdaoui et al.~\cite{BalabdaouiDurotETAL2019}. They assume that the true regression function is of the form $\E[Y|\bX = \bx] = \psi(\balpha^\top\bx)$, with an increasing function $\psi$. In contrast to our proposal, the regression model $\pi$ is fixed to be a linear model $\balpha^\top\bx$ in their approach. They consider global least squares estimation jointly for $(\psi,\balpha)$, but find it computationally intensive. As an alternative they suggest a two-step estimation procedure similar to our approach but with a split of the data such that $\balpha$ and the isotonic regression are estimated on independent samples. They find that if the rate of convergence of the estimator for $\balpha$ is sufficiently fast, then the resulting estimator of the true regression function is consistent with a 
convergence rate of order $n^{1/3}$. 

In a distributional regression framework, Henzi et al.~\cite{HenziKlegerETAL2021a} considered the described two-step estimation procedure with an isotonic distributional regression \cite{HenziZiegelETAL2021}, instead of a classical least squares isotonic regression as described in Section \ref{sec:isoreg}. They show that in both cases, with and without sample splitting, the procedure leads to consistent estimation of the conditional distribution of $Y$ given $\bX$, as long as the index $\pi$ can be estimated at a parametric rate. The two options, with and without sample splitting, do not result in relevant differences in predictive performance in the applications considered by Henzi et al.~\cite{HenziKlegerETAL2021a}. 


Assumption \eqref{eq:iso-assumption} can be checked by diagnostic plots using binning similarly to the plots in Henzi et al.~\cite[Figure 2]{HenziKlegerETAL2021a} in the distributional regression case. Predictive performance should be assessed on a test set of data disjoint from $(y_i,\bx_i)_{i=1}^n$, that is, on data that has not been used in the estimation procedure at all. Isotonic recalibration insures auto-calibration in-sample, and under an i.i.d.~assumption, auto-calibration will also hold approximately out-of-sample. Out-of-sample auto-calibration can be diagnosed with CORP (consistent, optimally binned, reproducible and PAV) mean reliability diagrams as suggested by Gneiting-Resin \cite{GneitingResin2021}, and comparison of predictive performance can be done with the usual squared error loss function or deviance loss functions.

\subsection{Over-fitting at the boundary}
\label{section over-fitting}
There is a small issue with the isotonic recalibration, namely, it tends to over-fit at the lower and upper boundaries
of the ranks $\widehat{\pi}(\bx_1) < \ldots < \widehat{\pi}(\bx_n)$. For instance, if $y_n$ is the largest observation in the portfolio (which is
not unlikely since the ranking $\widehat{\pi}$ is chosen response data-driven), then we estimate $\widehat{\mu}_{i_K}=y_n$, where $K = K((y_i,\widehat{\pi}(\bx_i))_{i=1}^n)$. Often, this over-fits
to the (smallest and largest) observations, as such extreme values/estimates cannot be verified on out-of-sample data. For this reason, we visually analyze the largest
and smallest values in the estimates $\widehat{\bmu}$, and we may manually merge, say, the smallest block ${\cal I}_1$ with
the second smallest one ${\cal I}_2$ (with the resulting estimate \eqref{PAVA block estimate} on the merged block).
More rigorously, this pooling could be cross-validated on out-of-sample data, but we refrain from doing so.
We come back to this in Figure \ref{Example 2}, below, where we merge the two blocks with the biggest estimates.

\subsection{Interpretation}
\label{sec:diagnostics} 

In \eqref{discrete interval} we have introduced the complexity number $K((y_i,\widehat{\pi}(\bx_i))_{i=1}^n)$ that counts the number of different values in $\widehat{\bmu}$, obtained by the isotonic regression \eqref{isotonic regression} in the isotonic recalibration step.
This complexity number $K((y_i,\widehat{\pi}(\bx_i))_{i=1}^n)$ allows one to assess the information content of the model, or in other words, how much signal is explainable from the data. Theorem \ref{signal-to-noise theorem} shows that the lower the signal-to-noise ratio, the lower the complexity number of the isotonic regression that we can expect. Clearly, in Theorem \ref{signal-to-noise theorem} we assume that the ranking of the observations is correct which will only be approximately satisfied since $\pi$ has to be estimated. In general, having large samples and flexible regression models for modeling $\pi$, it is reasonable to assume that the statement remains qualitatively valid. However, in complex (algorithmic) regression models, we need to ensure that we prevent from in-sample overfitting; this is typically controlled by either using (independent) validation data or by performing a cross-validation analysis.

Typical claims data in non-life insurance have a low signal-to-noise ratio. Regarding claims frequencies, this low signal-to-noise ratio is caused by the fact that claims are not very frequent events, e.g., in car insurance annual claims frequencies range from 5\% to 10\%, that is, only one out of 10 (or 20) drivers suffers a claim within a calendar year. A low signal-to-noise ratio also applies to claim amounts, which are usually strongly driven by randomness and the explanatory part from policyholder information is comparably limited. Therefore, we typically expect a low complexity number $K((y_i,\widehat{\pi}(\bx_i))_{i=1}^n)$ both for claims frequency and claim amounts modeling.

In case of a small to moderate complexity number $K=K((y_i,\widehat{\pi}(\bx_i))_{i=1}^n)$, the regression function $\widehat{\bmu}$ becomes interpretable through the isotonic recalibration step. For this, we extend the auto-calibrated regression function $\widehat{\mu}$ from the set $\{\widehat{\pi}(\bx_1), \dots,\widehat{\pi}(\bx_n)\}$ to the entire covariate space $\mathcal{X}$ by defining a step function 
\[
\widehat{\mu}(\bx) = \widehat{\mu}_{i_k}, \quad \text{if \quad $\widehat{\pi}(\bx_{i_{k}}) \le \widehat{\pi}(\bx) < \widehat{\pi}(\bx_{i_{k+1}})$},
\]
for all $\bx \in \mathcal{X}$, where $0=i_0 < i_1 < \dots < i_K=n$ are the slicing points of the isotonic regression as defined in \eqref{discrete interval}. Figure \ref{fig:simpleiso} illustrates this step function interpolation which is different from an interpolation scheme that one would naturally use for prediction. We define a partition $\mathcal{X}_{1},\dots,\mathcal{X}_{K}$ of the original covariate space ${\cal X}$ by
\begin{equation}\label{eq:def_partition}
\mathcal{X}_{k} = \{\bx \in \mathcal{X} : \widehat{\mu}(\bx) = \widehat{\mu}_{i_k}\}, \quad k = 1,\dots,K.
\end{equation}
Figure \ref{Example 1 0} illustrates how this partition of $\mathcal{X}$ provides insights on the covariate-response relationships in the model. This procedure has some analogy to regression trees and boosting trees that rely on partitions of the covariate space ${\cal X}$. In the case study in Section \ref{sec:examples}, we illustrate two further possibilities to use the partition defined at \eqref{eq:def_partition} for understanding covariate-response relationships. First, in Figure \ref{Marginal plots Example 2}, the influence of individual covariates on the price cohorts is analyzed, and second, Figure \ref{Feature space partition} gives a summary view of the whole covariate space for a chosen price cohort.

\section{Swedish motorcycle data}\label{sec:examples}

We consider claim amounts modeling on the Swedish motorcycle data which was originally presented in
the text book of Ohlsson--Johansson \cite{Ohlsson} and which is also studied in W\"uthrich--Merz \cite{WM2023}.\footnote{The Swedish motorcycle data set is available through the {\sf R} package {\tt CASdatasets} \cite{DutangCharpentier}.}
This data set comprises comprehensive insurance for motorcycles in Sweden. The insurance product covers loss or damage of
motorcycles other than collision, e.g., caused by theft, fire or vandalism. The data contains claims aggregated
per feature (covariate) combination for the calendar years 1994–1998. There are 683 claims on 62,036 different covariates, thus, 
claims are very sparse.  
We use exactly the same data pre-processing as described in \cite[Listing 13.3]{WM2023},
and an excerpt of the pre-processed data is shown in Listing \ref{DataSweden}; for a description of the different covariates
we refer to \cite[Section 2.4]{Ohlsson} and  \cite[Section 13.2]{WM2023}. The goal is to build a regression model for these 683 positive claim amounts, and use isotonic recalibration for 
auto-calibration and interpretation as described in Section \ref{sec:diagnostics}.

\begin{lstlisting}[float=h,frame=tb,caption={Excerpt of the Swedish motorcycle data set.},
label=DataSweden]
'data.frame':   62036 obs. of  9 variables:
 $ OwnerAge   : num  18 18 18 18 18 18 18 18 18 18 ...
 $ Gender     : Factor w/ 2 levels "Female","Male": 1 1 1 1 1 1 1 1 1 1 ...
 $ Area       : Factor w/ 7 levels "Zone 1","Zone 2",..: 1 1 1 1 2 2 2 3 3 3 ...
 $ RiskClass  : int  1 2 3 3 1 1 3 1 1 1 ...
 $ VehAge     : num  8 11 9 9 11 12 24 4 6 6 ...
 $ BonusClass : int  2 2 3 4 1 1 2 1 1 2 ...
 $ Exposure   : num  1 0.778 0.499 0.501 0.929 ...
 $ ClaimNb    : int  0 0 0 0 0 0 0 0 0 0 ...
 $ ClaimAmount: int  0 0 0 0 0 0 0 0 0 0 ...
\end{lstlisting}

\subsection{Isotonic recalibration vs.~binary regression trees}
We start by considering the two covariate components {\tt RiskClass} and {\tt VehAge} only. Since the resulting covariate space ${\cal X} =\{({\tt RiskClass}, {\tt VehAge})\}\subset \R^2$ is two-dimensional, we can graphically illustrate the differences between the isotonic recalibration approach and a binary regression tree (as a competing model) for interpretation. In Section \ref{sec:allcovariates}, we consider all available covariates.

We fit a deep feed-forward neural network (FFNN) regression model to these 683 claims. We choose a network architecture of depth 3 with $(20,15,10)$ neurons in the three hidden layers, the hyperbolic tangent activation function in the hidden layers, and the log-link for the output layer. The input has dimension 2, this results in a FFNN architecture with
a network parameter of dimension 546; for a more detailed discussion of FFNNs we refer to \cite[Chapter 7]{WM2023}, in particular, to
Listings 7.1-7.3 of that reference. We fit this model using the gamma deviance loss, see \cite[Section 5.3.7]{WM2023} and Remark \ref{rem:Bregman},
use the {\tt nadam} version of stochastic gradient descent, and exercise early stopping on a validation set being
20\% of the entire data. 
Line (1a) of Table \ref{results gamma 0}, called gamma FFNN, shows the performance of the fitted FFNN regression model.  This is compared
to the null model (empirical mean) on line (0) that does not consider any covariates.\footnote{In a gamma null model, i.e., assuming i.i.d.~gamma distributed responses, we obtain that the MLE of the mean is equal to the empirical mean of the observations; this generally holds true within the exponential dispersion family.} We observe a decrease in gamma deviance loss and in root mean squared error (RMSE) which justifies the use of a regression model; note that these
are in-sample figures, but we use early stopping to prevent the network from in-sample overfitting. The difficulty here is that, only having 683 claims, we cannot provide a reasonable out-of-sample analysis. The last column of Table \ref{results gamma 0} called 'average' compares the average claims estimate of the FFNN to the empirical mean, and we observe a slight positive bias in the FFNN prediction, i.e., $24,932>24,641$.

\begin{table}[htb!]
\centering
{\small
\begin{center}
\begin{tabular}{|cl||cc|c|}
\hline
& &gamma deviance &RMSE &average\\
\hline
(0)&  null model &2.085& 35,311&24,641\\\hline
(1a)&  gamma FFNN &1.704& 32,562&24,932\\
  (1b)&  gamma FFNN recalibrated &1.640& 32,005&24,641\\\hline
  (2)&  binary regression tree &1.761& 32,706&24,641\\\hline
\end{tabular}
\end{center}}
\caption{Loss figures in the Swedish motorcycle example only considering {\tt RiskClass} and {\tt VehAge} as covariates. }
\label{results gamma 0}
\end{table}

In the next step, we use the FFNN estimates as ranks $\widehat{\pi}(\bx_i)$ for ordering the claims $y_i$ and the covariates
$\bx_i$, respectively. Then we apply the non-parametric isotonic recalibration step \eqref{isotonic regression} to these
ranks and claims. The Swedish motorcycle claims data is aggregated w.r.t.~the available covariate combinations, and the
683 positive claims come from 656 different covariate combinations $\bx_i$. This requires that we work with the weighted
version of \eqref{isotonic regression}, where $w_i \in \N$ corresponds to the number of claims that have been
observed for covariate $\bx_i$, and $y_i$ corresponds to the average observed claim amount on $\bx_i$.\footnote{Since
  we only consider the two covariate components {\tt RiskClass} and {\tt VehAge} in this example, we further aggregate the claims over
  these covariate combinations. This results in sufficient statistics for the gamma regression model, and we only need
  to adjust the weights $w_i$ correspondingly. This is an elegant way of avoiding to deal with ties for continuous
regression functions (and supposed that the aggregation within different covariate combinations is computationally feasible).}
We use the {\sf R} package {\tt monotone} \cite{monotone} which provides a fast implementation of the PAV algorithm.
The numerical results are presented on line (1b) of Table \ref{results gamma 0}. There is a slight decrease in average loss through the isotonic recalibration. This is expected since the isotonic regression is optimizing the in-sample loss for any Bregman loss function, see Remark \ref{rem:Bregman}. 
The last column of Table \ref{results gamma 0}
verifies that now the global balance property \eqref{global balance} holds.

\begin{figure}[htb!]
\begin{center}
\begin{minipage}[t]{0.4\textwidth}
\begin{center}
\includegraphics[width=\textwidth]{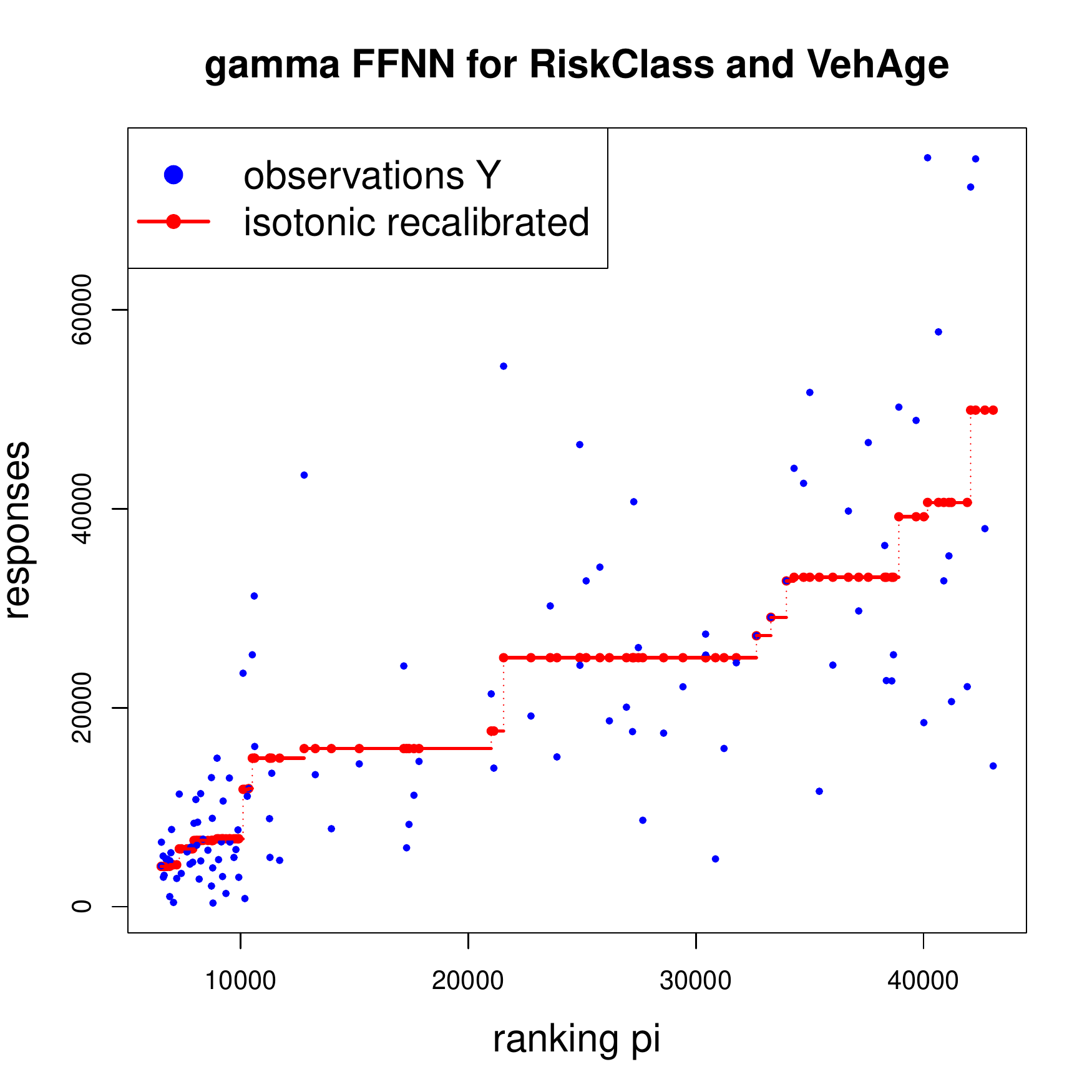}
\end{center}
\end{minipage}
\end{center}
\vspace{-.7cm}
\caption{Isotonic recalibration in the Swedish motorcycle example only using
  {\tt RiskClass} and {\tt VehAge} as covariates resulting in the complexity
  number $K((y_i,\widehat{\pi}(\bx_i))_{i=1}^n)=18$.}
\label{Example 1}
\end{figure}

Figure \ref{Example 1} provides the resulting step function from the isotonic recalibration (in red color) of the ranking
$(\widehat{\pi}(\bx_i))_{i=1}^n$ given by the gamma FFNN; this is complemented
with the observed amounts $y_i$ (in blue color). The resulting complexity number is $K=K((y_i,\widehat{\pi}(\bx_i))_{i=1}^n)=18$, i.e., in this
example the conditional expected claim amounts can be represented by 18 different estimates $\widehat{\mu}_{i_k} \in \R$, $k=1,\ldots, K=18$;
the FFNN regression function uses $6\cdot 21=126$ different values (ranks) which corresponds to the
cardinality of the available covariate values $({\tt RiskClass},{\tt VehAge}) \in {\cal X}$.

\begin{figure}[htb!]
\begin{center}
\begin{minipage}[t]{0.4\textwidth}
\begin{center}
\includegraphics[width=\textwidth]{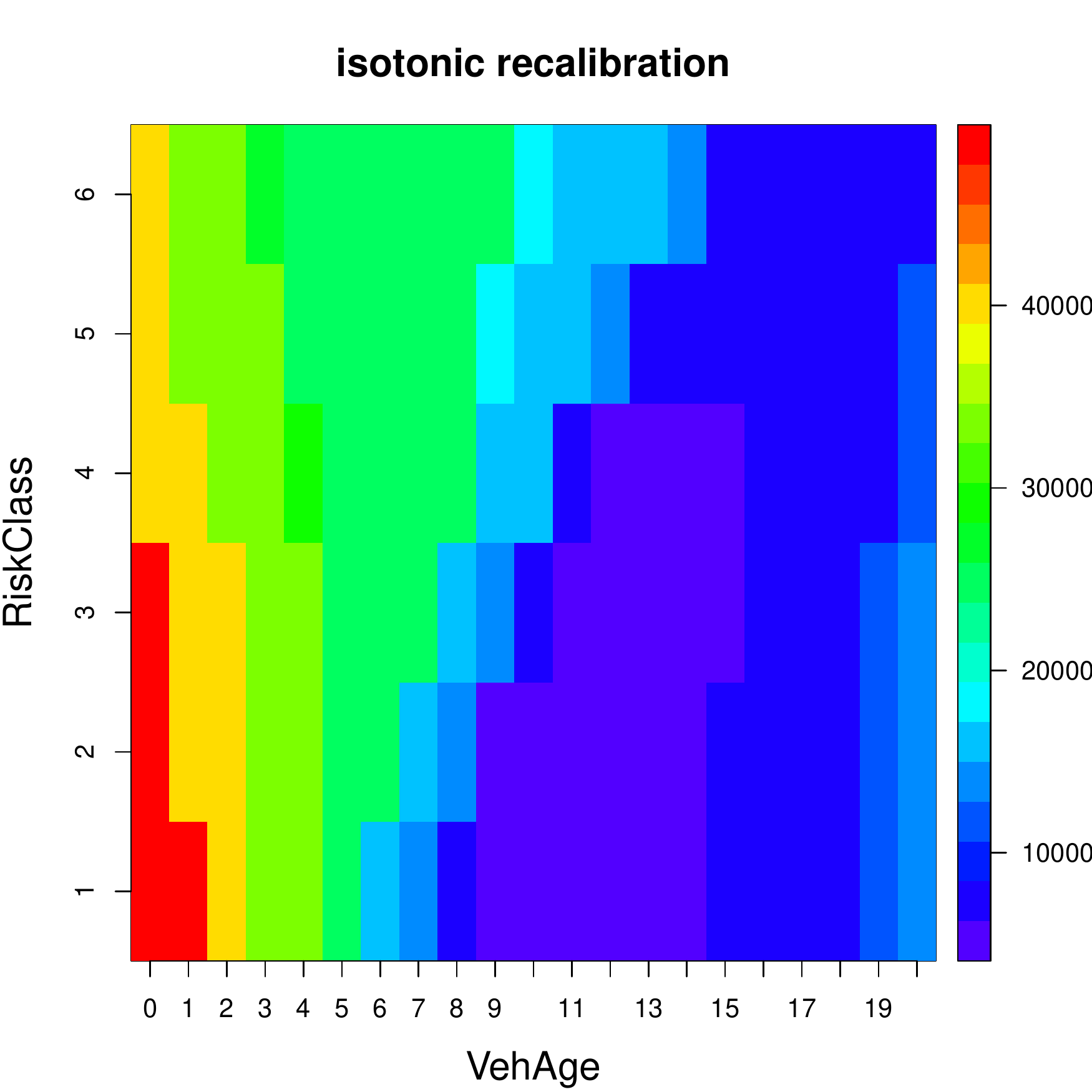}
\end{center}
\end{minipage}
\begin{minipage}[t]{0.02\textwidth}
\begin{center}
~
\end{center}
\end{minipage}
\begin{minipage}[t]{0.4\textwidth}
\begin{center}
\includegraphics[width=\textwidth]{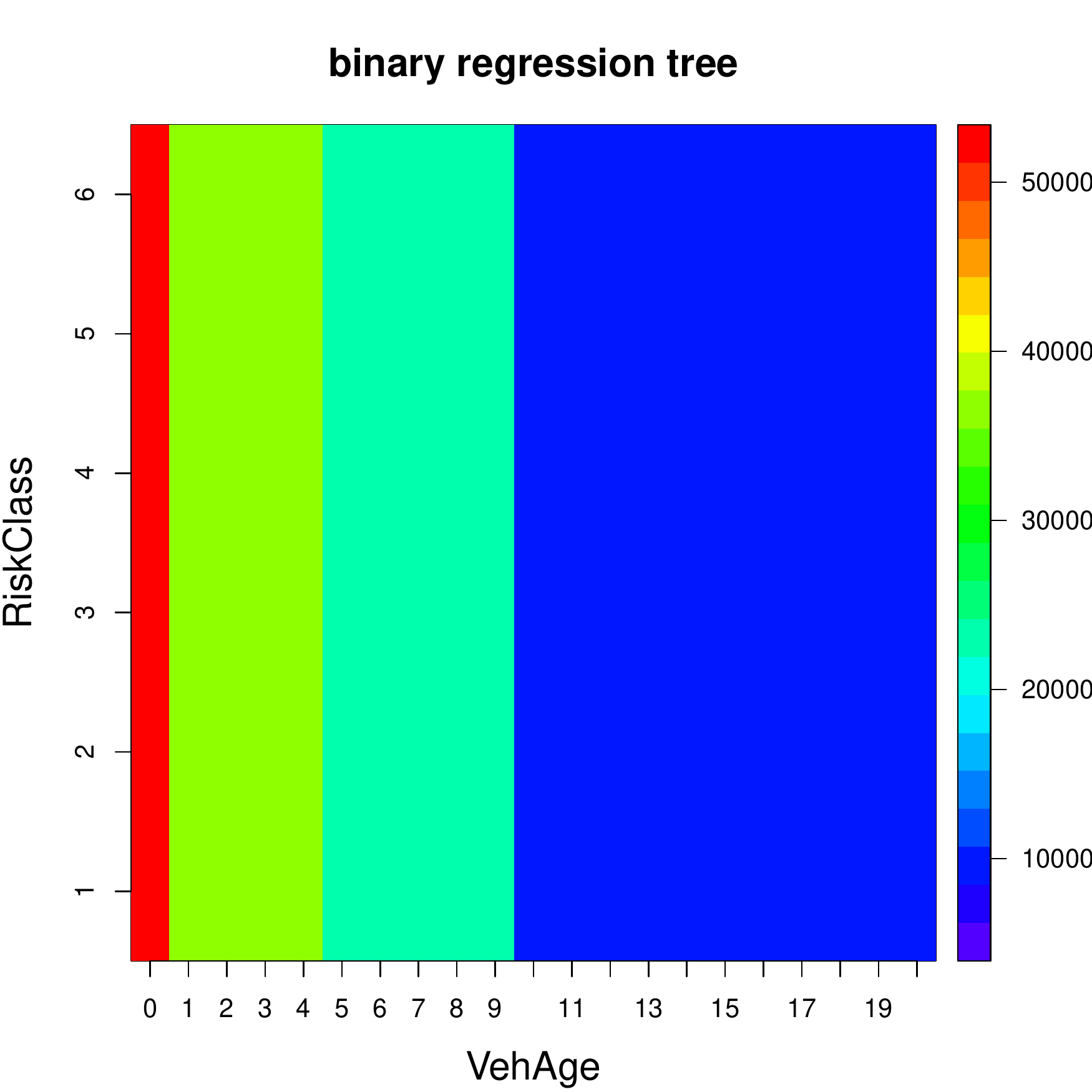}
\end{center}
\end{minipage}
\end{center}
\vspace{-.7cm}
\caption{(lhs) Isotonic recalibration and (rhs) binary regression tree, both only using
  {\tt RiskClass} and {\tt VehAge} as covariates; the color scale is the same in both plots.}
\label{Example 1 0}
\end{figure}

The isotonic recalibration on the ranks $\widehat{\pi}(\bx) = \widehat{\pi}({\tt RiskClass},{\tt VehAge})$ of the FFNN
leads to a partition $\mathcal{X}_1,\dots,\mathcal{X}_{18}$ of the
covariate space as defined at \eqref{eq:def_partition}. 
We compare this partition to the one that results  from a binary split regression tree
approach. We use 10-fold cross-validation to determine the optimal tree size. In this example the
optimal tree has only 3 splits, and they all concern the variable {\tt VehAge}. The resulting
losses of this optimal tree are shown on line (2) of Table \ref{results gamma 0}, and we conclude that the regression tree
approach is not fully competitive, here. More interestingly, Figure \ref{Example 1 0} shows the
resulting partitions of the covariate space ${\cal X}=\{({\tt RiskClass},{\tt VehAge})\}$ from the two
approaches. The plot on the right-hand side shows the three splits of the regression tree (all w.r.t.~{\tt VehAge}).  From the
isotonic recalibration approach on the left-hand side, we learn that a good regression model should have
diagonal structures, emphasizing that the two covariates interact in a nontrivial way which cannot be captured by the binary split
regression tree in this case.

\subsection{Consideration of all covariates}\label{sec:allcovariates}
We now consider all available covariate components, see lines 2-7 of Listing \ref{DataSweden}. We first fit a FFNN to this
data. This is done exacly as in the previous example with the only difference that the input dimension changes from 2 to 6, when we consider
all available information.  We transform the (ordered) {\tt Area} code into  real values, and also we also merge {\tt
  Area} codes 5 to 7 because of scarcity of claims for these {\tt Area} codes, and we call this new
variable {\tt Zone}. The FFNN has then a network parameter of dimension 626.
The network is fitted with stochastic gradient descent that is early stopped based on a validation loss analysis.
The results are presented on line (2a) of Table \ref{results gamma 1}.

\begin{table}[htb!]
\centering
{\small
\begin{center}
\begin{tabular}{|cl||cc|c|}
\hline
& &gamma deviance &RMSE &average\\
\hline
  (0)&  null model &2.085& 35,311&24,641\\\hline
  (1a)&  gamma GLM &1.717& 32,562&25,105\\  
   (1b)&  gamma GLM recalibrated with $K=24$&1.641& 31,578&24,641\\\hline  
(2a)&  gamma FFNN &1.496& 29,673&24,526\\
  (2b)&  gamma FFNN recalibrated with $K=22$ &1.452& 28,806&24,641\\
  (2c)&  gamma FFNN tree adjustment with 4 bins (seed 1) &1.508& 29,371&24,641\\
  (2d)&  gamma FFNN tree adjustment with 8 bins (seed 2) &1.466& 27,942&24,641\\\hline
\end{tabular}
\end{center}}
\caption{Losses in the Swedish motorcycle example based on all available covariates. }
\label{results gamma 1}
\end{table}

We compare the fitted FFNN regression model to the null model (empirical mean) and a gamma generalized linear model (GLM).
The gamma GLM is identical to model Gamma GLM1 in \cite[Table 5.13]{WM2023}. We give some remarks on the results of Table
\ref{results gamma 1}. Firstly, the FFNN has the smallest gamma deviance loss and the smallest RMSE
of the three models on lines (0)-(2a). Thus, the gamma FFNN adapts best to the data among the three model choices (we use early stopping in the FFNN fitting).
Interestingly, the gamma GLM
and the FFNN both fail to have the global balance property \eqref{global balance}, see last column of Table \ref{results gamma 1}. Stochastic
gradient descent fitted models with early stopping generally fail to satisfy the global balance property, whereas the gamma GLM
fails to have the global balance property because we work with the log-link and not with the canonical link of the Gamma GLM, here.

\begin{figure}[htb!]
\begin{center}
\begin{minipage}[t]{0.32\textwidth}
\begin{center}
\includegraphics[width=\textwidth]{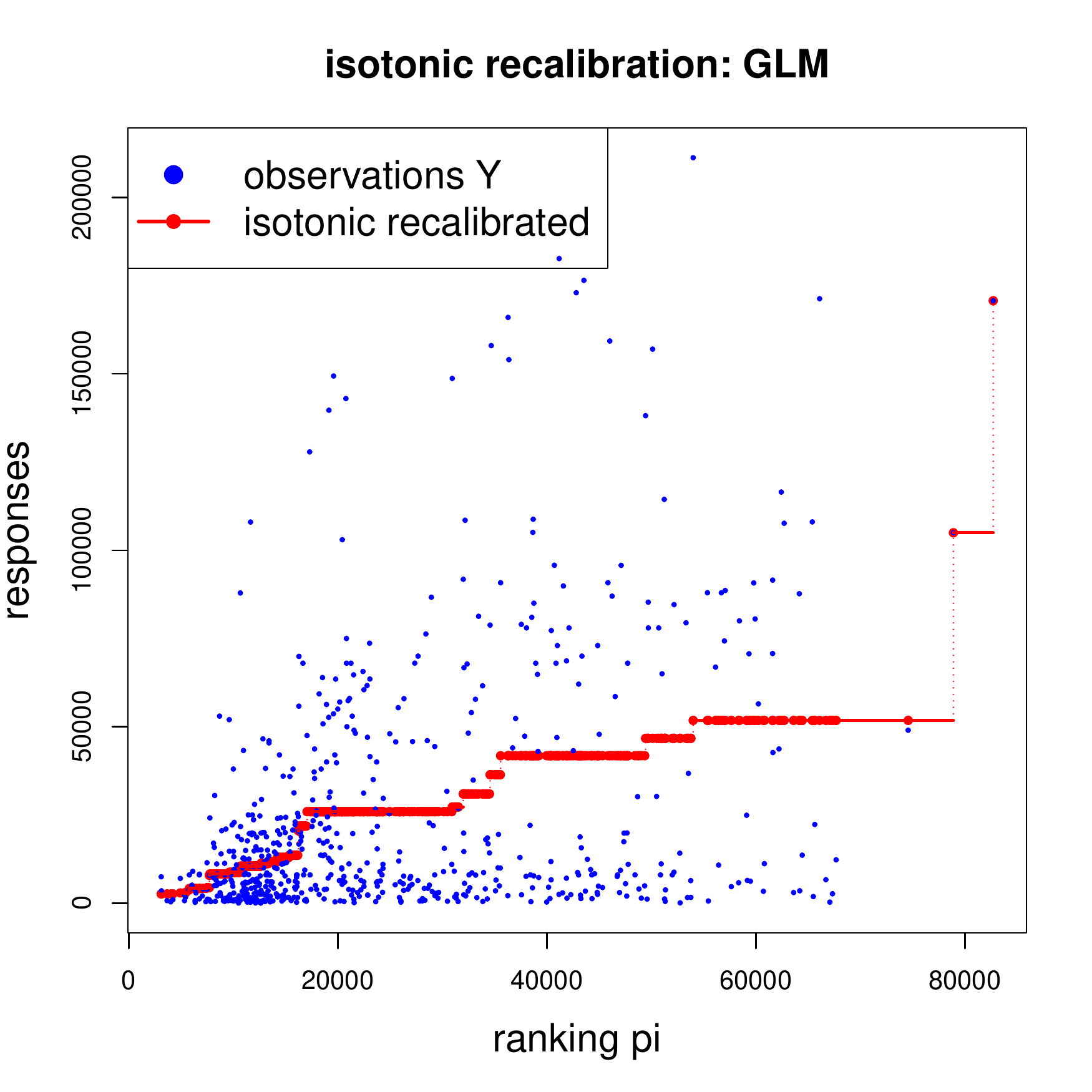}
\end{center}
\end{minipage}
\begin{minipage}[t]{0.32\textwidth}
\begin{center}
\includegraphics[width=\textwidth]{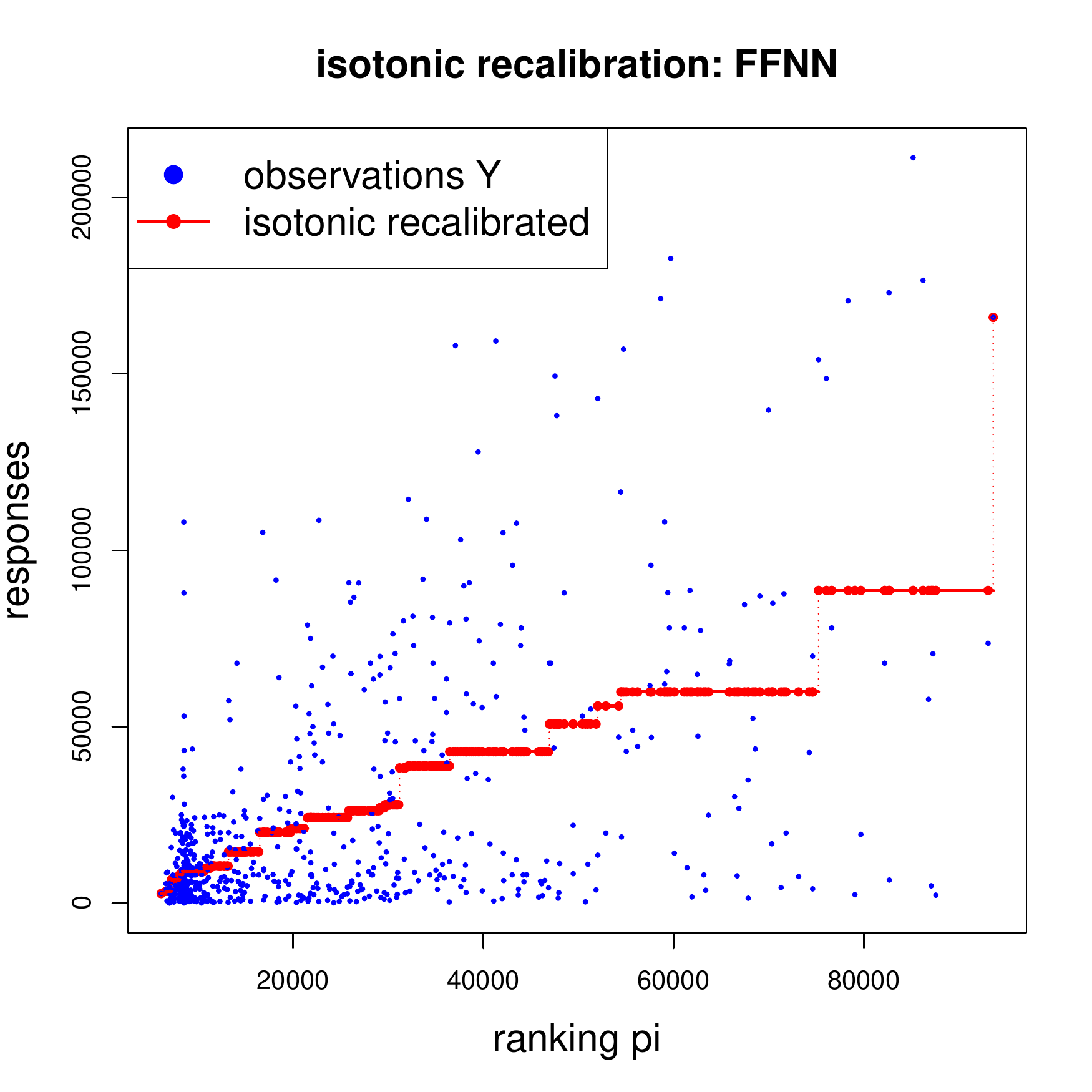}
\end{center}
\end{minipage}
\begin{minipage}[t]{0.32\textwidth}
\begin{center}
\includegraphics[width=\textwidth]{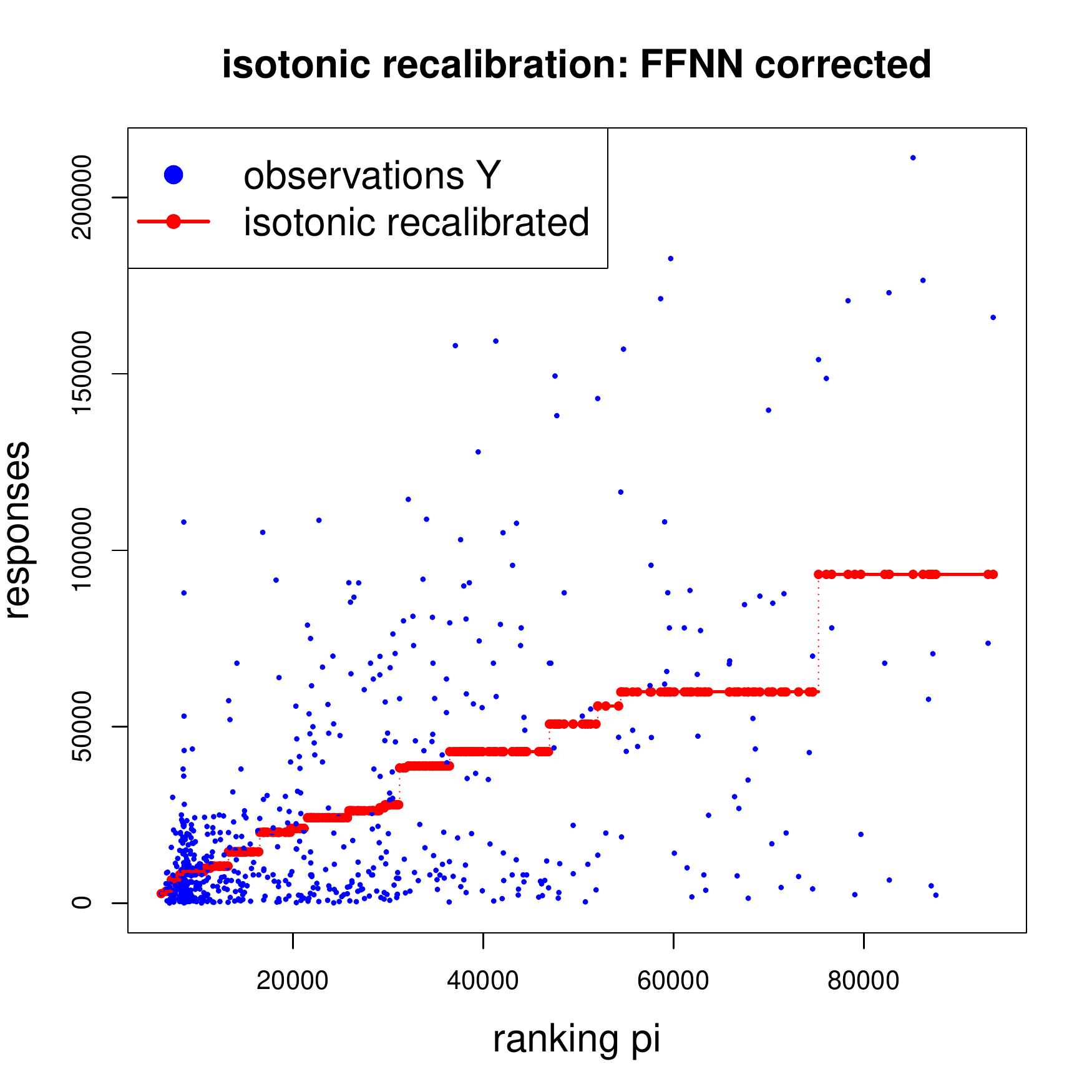}
\end{center}
\end{minipage}
\end{center}
\vspace{-.7cm}
\caption{Isotonically recalibrated regression models in the Swedish motorcycle example using
  all covariates for the gamma GLM with complexity number $K((y_i,\widehat{\pi}(\bx_i))_{i=1}^n)=24$ (lhs), for the gamma FFNN
  with complexity number $K((y_i,\widehat{\pi}(\bx_i))_{i=1}^n)=23$ (middle) and over-fitting corrected (rhs).}
\label{Example 2}
\end{figure}

In the next step, we use the FFNN predictions as ranks $\widehat{\pi}(\bx_i)$ for ordering the responses and covariates,
and we label the claims $y_i$ such that $\widehat{\pi}(\bx_1)<\ldots <\widehat{\pi}(\bx_n)$. There are no ties in this data, and we obtain
$n=656$ pairwise different values. The results of the isotonic recalibration are presented in Figure \ref{Example 2} (middle).
The complexity number is $K=K((y_i,\widehat{\pi}(\bx_i))_{i=1}^n)=23$, thus, the entire regression problem is encoded in 23 different
values $\widehat{\mu}_{i_k}$, $k=1,\dots, K$. In view of this plot, it seems that the largest value 
$\widehat{\mu}_{i_K}$ over-fits to the corresponding observation, as this estimate is determine by a single observation $y_n$,
being bigger than the weighted block mean $\widehat{\mu}_{i_{K-1}}$ on the previous block ${\cal I}_{K-1}$; compare
Section \ref{section over-fitting}. For this reason, we manually pool the two last blocks ${\cal I}_{K-1}$
and ${\cal I}_{K}$. This provides us with a new estimate \eqref{PAVA block estimate} on this merged block,
and reduces the complexity number by 1 to $K=22$. The resulting isotonic recalibration is shown in Figure
\ref{Example 2} (rhs), and the empirical losses are provided on line (2b) of Table \ref{results gamma 1}.
Importantly, this isotonic recalibrated regression is in-sample auto-calibrated \eqref{auto-calibration step 2} and, henceforth,
it fulfills the global balance property which can be verified in the last column of Table \ref{results gamma 1}.

We perform the same isotonic recalibration to the ranks obtained from the gamma GLM in Table \ref{results gamma 1}. We observe that the isotonic recalibration step leads to a major decrease in average loss in the gamma GLM, and it results in the complexity number $K=24$, see
also Figure \ref{Example 2} (lhs).

We compare isotonic recalibration to a recent proposal of Lindholm et al.~\cite{Lindholm} that also 
achieves auto-calibration in-sample. Isotonic regression provides a partition of the index set ${\cal I}=\{1,\dots,n\}$
into disjoint blocks ${\cal I}_1, \ldots, {\cal I}_K$ on which the estimated regression function is
constant. This can also be achieved by considering a binary regression tree algorithm applied to the (rank) covariates $\{\widehat{\pi}(\bx_i);\,1\le i \le n\}$ and corresponding responses $y_i$; see Section 2.3.2 of Lindholm et al.~\cite{Lindholm}. We call this latter approach the tree
binning approach. There are two main differences between the
tree binning approach and the isotonic recalibration approach. First, generally, the tree binning approach does not provide
a regression function that has the same ranking as the first regression step providing $\widehat{\pi}(\bx_i)$. Second, in the isotonic
regression approach, the complexity number $K((y_i,\widehat{\pi}(\bx_i))_{i=1}^n)$ is naturally given, i.e., the isotonic regression
\eqref{isotonic regression} automatically extracts the degree of information contained in the responses $\by$, and generally, this
degree of information is increasing for an increasing signal-to-noise ratio by Theorem \ref{signal-to-noise theorem}. Conversely, in the tree binning
approach, we need to determine the optimal number of bins (leaves), e.g., by $k$-fold cross-validation. The obtained number of bins depends on the
hyperparameters of the minimal leaf size and of the number of folds in cross-validation, as well as on the random
partition of the instances for cross-validation. We found that the number of bins is sensitive to the tuning choices, and hence, contrary to isotonic recalibration, the resulting partition is subject to potentially subjective choices and randomness.

\begin{figure}[htb!]
\begin{center}
\begin{minipage}[t]{0.4\textwidth}
\begin{center}
\includegraphics[width=\textwidth]{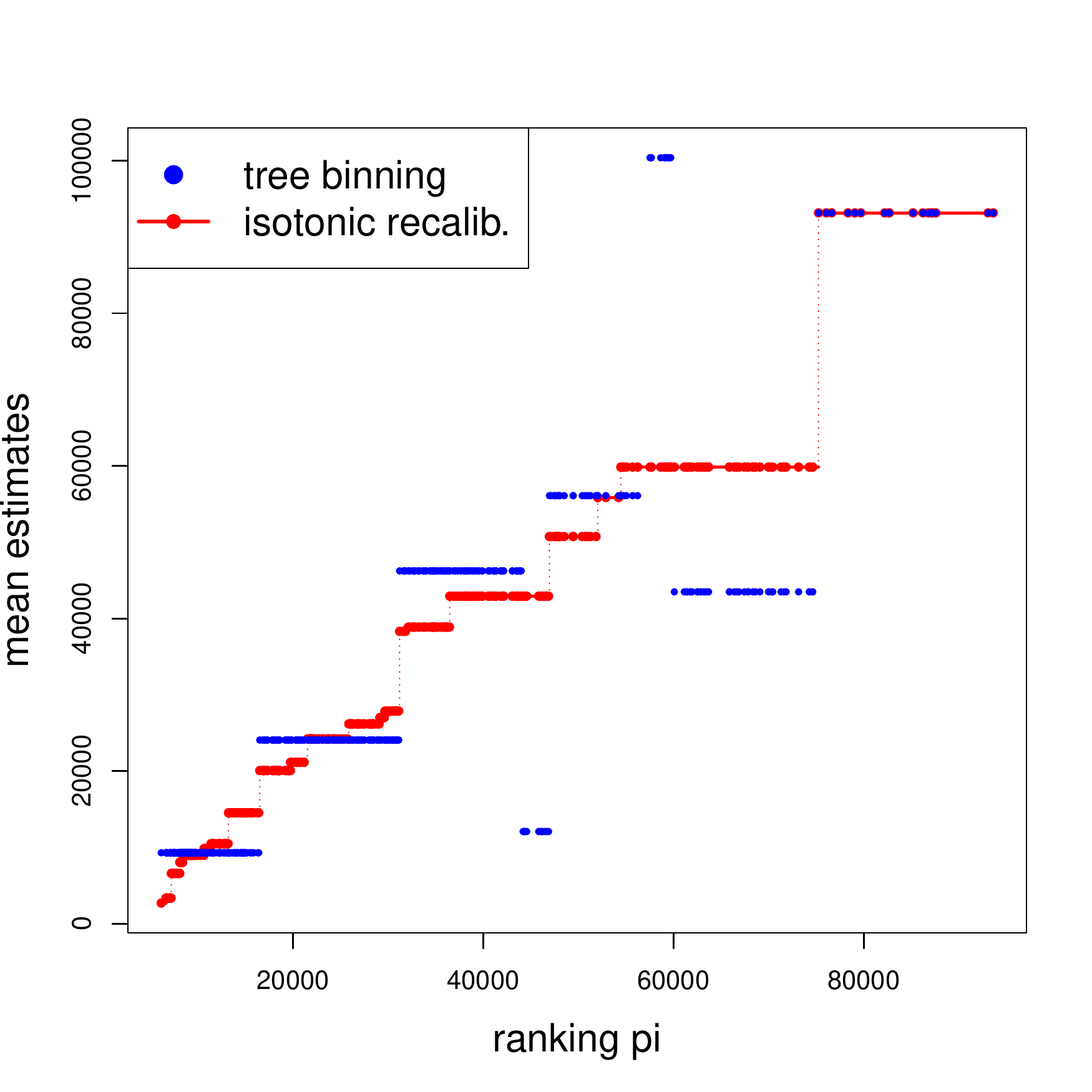}
\end{center}
\end{minipage}
\end{center}
\vspace{-.7cm}
\caption{Tree binning vs.~isotonic recalibration; the step functions correspond to lines
  (2d) and (2b) of Table \ref{results gamma 1} with 8 bins for line  (2d) and complexity number $K=22$
  for line (2b).}
\label{Example 2 binning}
\end{figure}

For the results on the tree binning approach in Table \ref{results gamma 1} we have chosen
$k=10$ folds and a minimal leaf size of $10$, and only the random partitioning of the pseudo-sample is different for the results
in lines (2c)-(2d). A first random seed gives 4 bins and a second one 8 bins, and we observe a considerable difference in the two models with respect to gamma deviance loss and the RMSE. 
Figure \ref{Example 2 binning} shows the isotonic recalibration and the tree binning approach with 8 bins, corresponding to lines (2b) and (2d)
of Table \ref{results gamma 1}. From this plot, we conclude that the tree binning approach does not necessarily preserve
the rankings induces by $\widehat{\pi}(\bx_i)$ as the resulting step function (in blue color) is not monotonically increasing.
We recommend isotonic recalibration to achieve auto-calibration since it preserves monotonicity of the regression model in the first estimation step, and there are no potentially influential tuning parameters.

\begin{figure}[htb!]
\begin{center}
\begin{minipage}[t]{0.31\textwidth}
\begin{center}
\includegraphics[width=\textwidth]{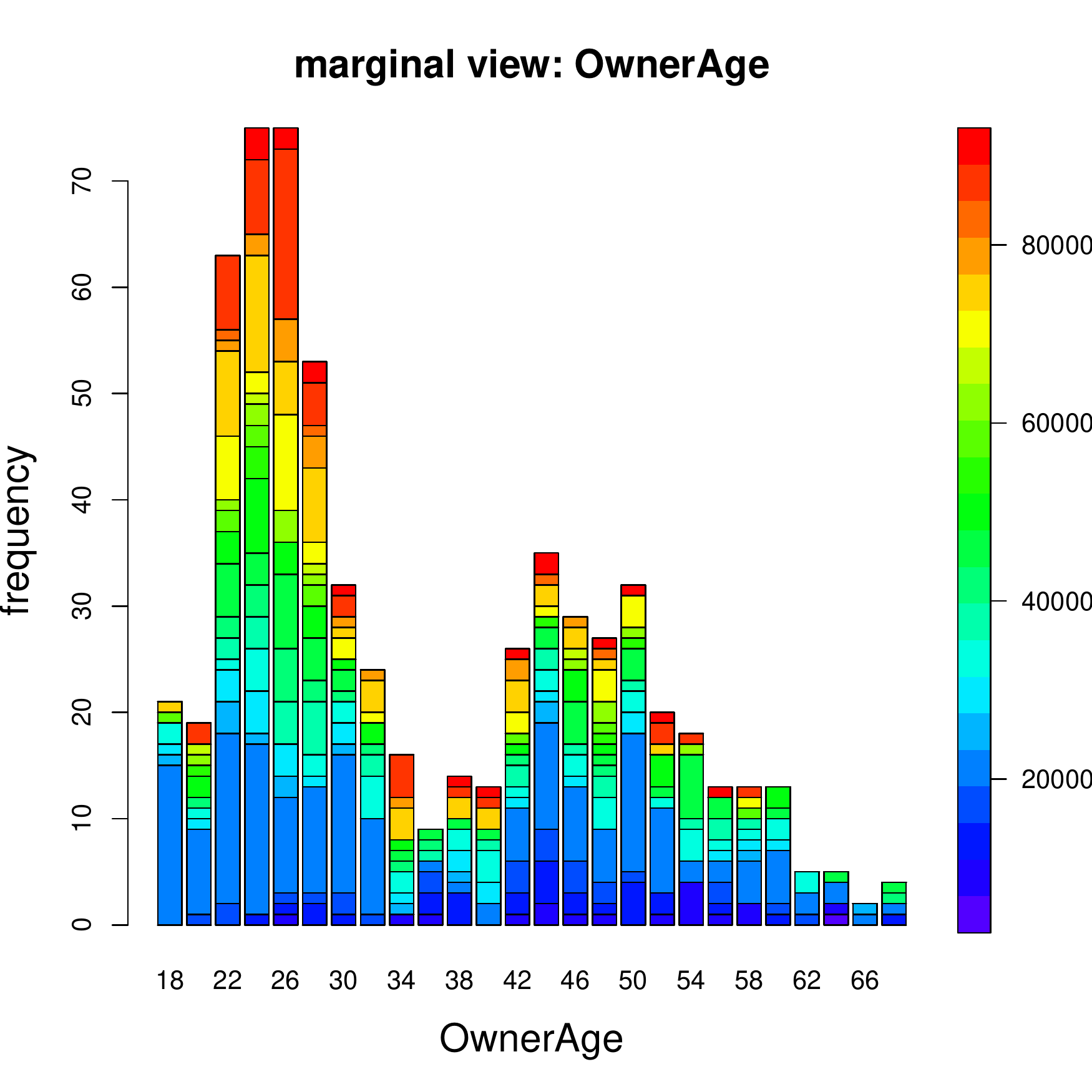}
\end{center}
\end{minipage}
\begin{minipage}[t]{0.02\textwidth}
\begin{center}
~
\end{center}
\end{minipage}
\begin{minipage}[t]{0.31\textwidth}
\begin{center}
\includegraphics[width=\textwidth]{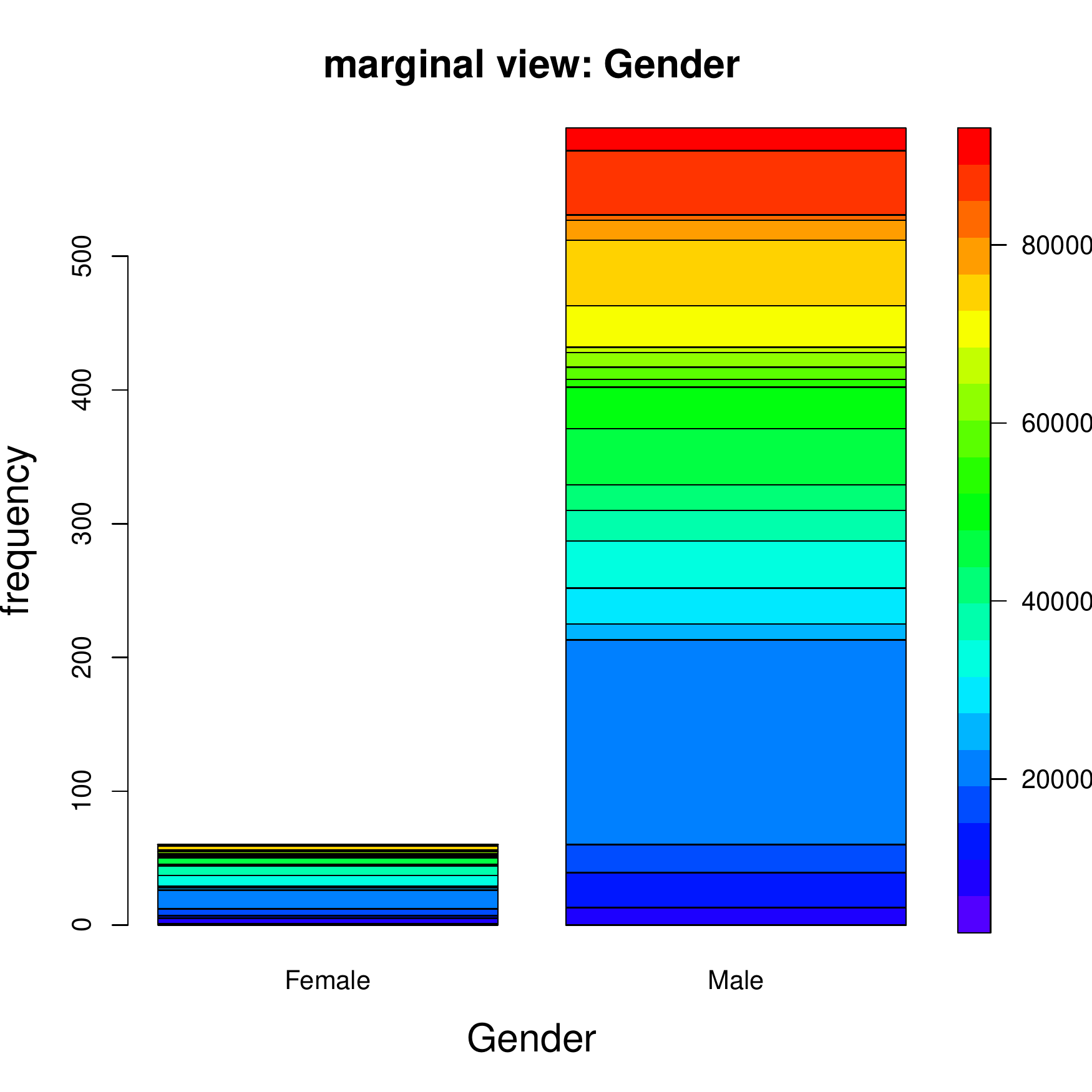}
\end{center}
\end{minipage}
\begin{minipage}[t]{0.02\textwidth}
\begin{center}
~
\end{center}
\end{minipage}
\begin{minipage}[t]{0.31\textwidth}
\begin{center}
\includegraphics[width=\textwidth]{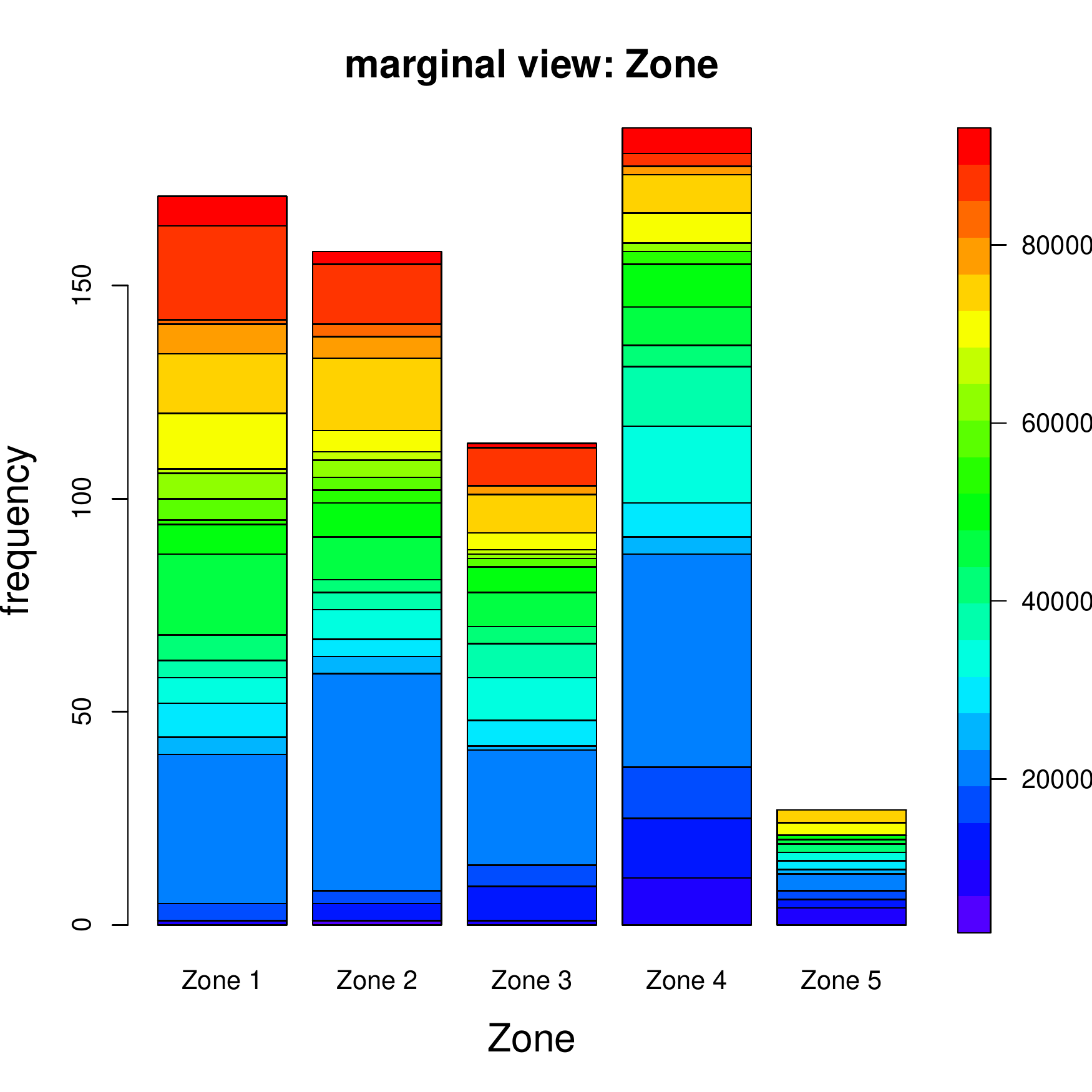}
\end{center}
\end{minipage}

\begin{minipage}[t]{0.31\textwidth}
\begin{center}
\includegraphics[width=\textwidth]{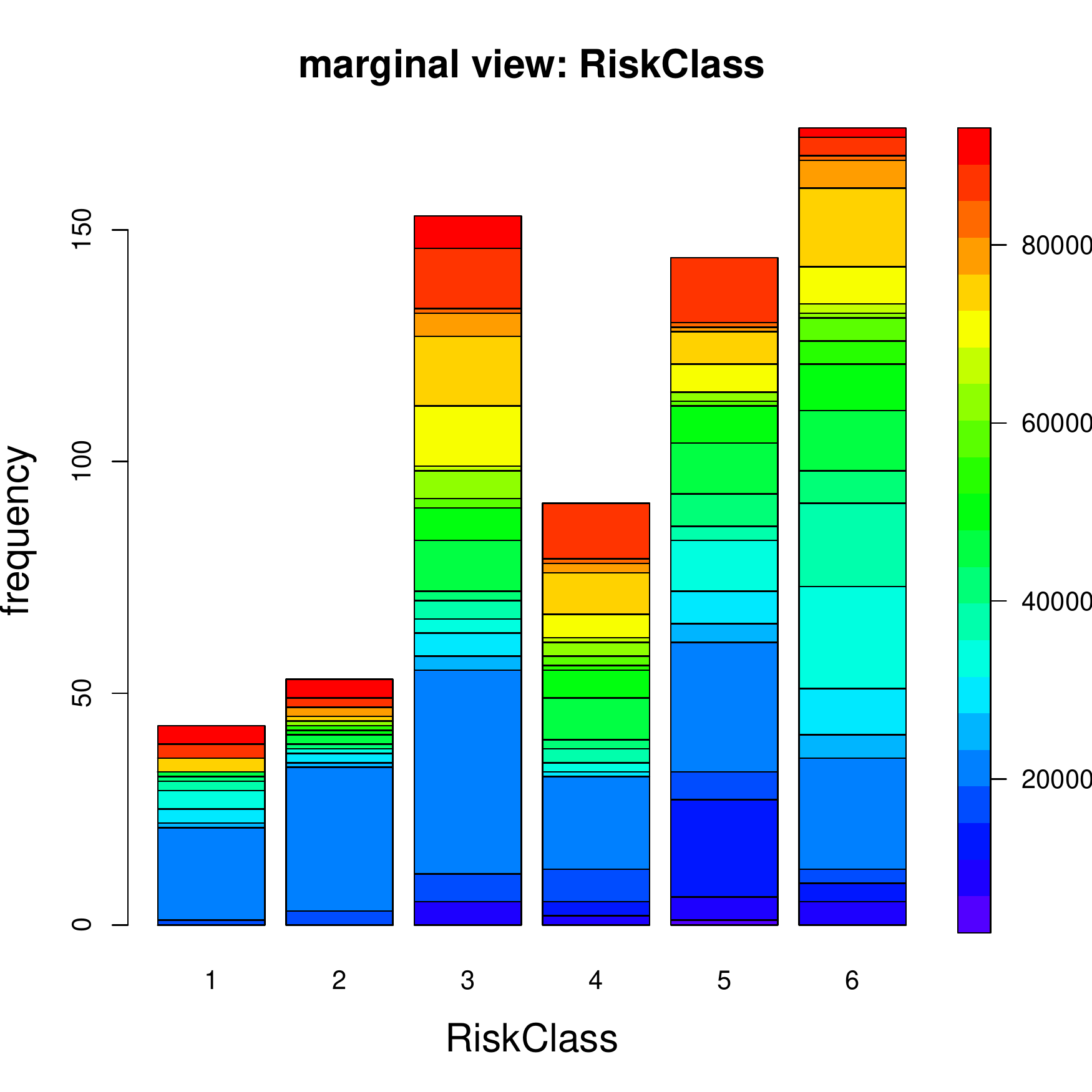}
\end{center}
\end{minipage}
\begin{minipage}[t]{0.02\textwidth}
\begin{center}
~
\end{center}
\end{minipage}
\begin{minipage}[t]{0.31\textwidth}
\begin{center}
\includegraphics[width=\textwidth]{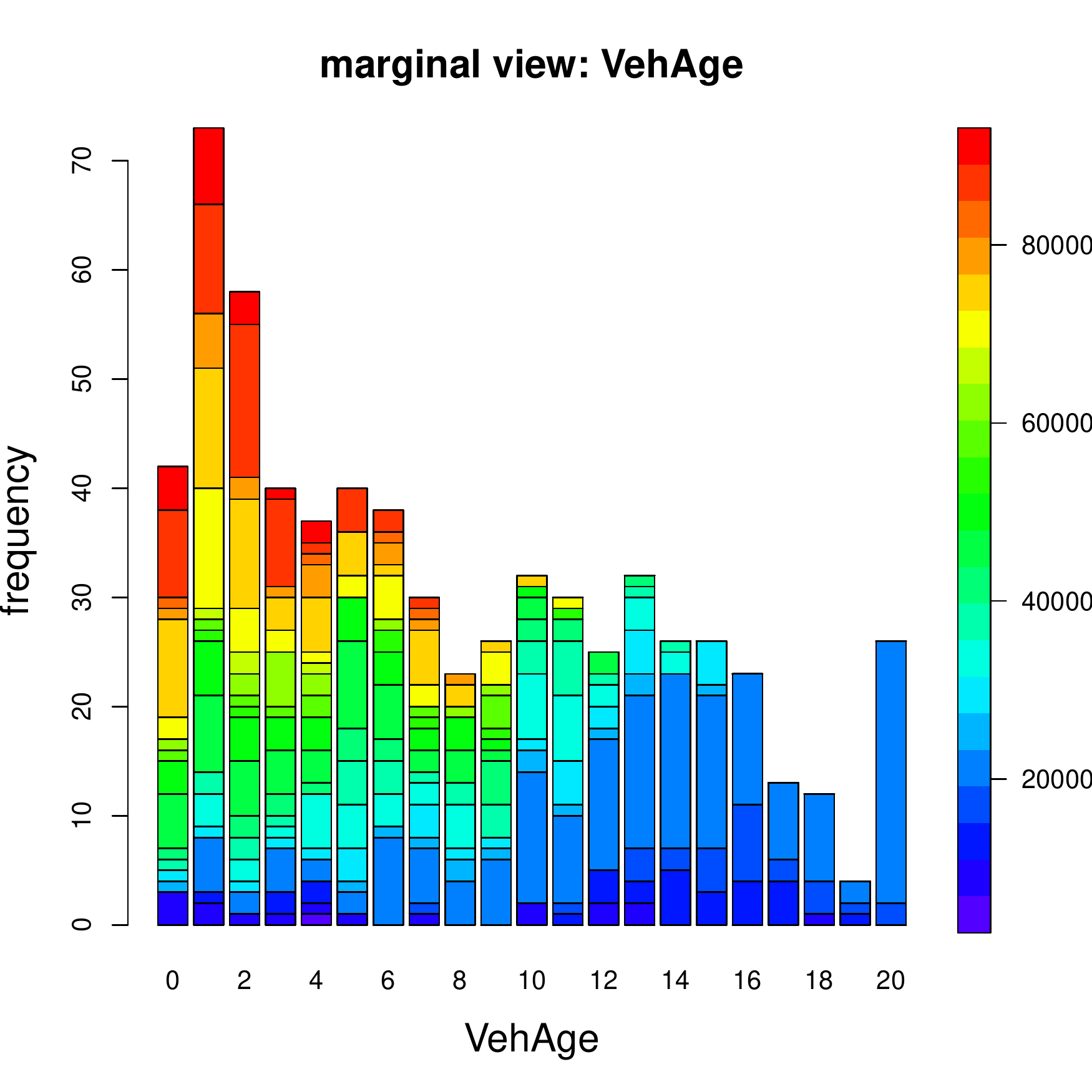}
\end{center}
\end{minipage}
\begin{minipage}[t]{0.02\textwidth}
\begin{center}
~
\end{center}
\end{minipage}
\begin{minipage}[t]{0.31\textwidth}
\begin{center}
\includegraphics[width=\textwidth]{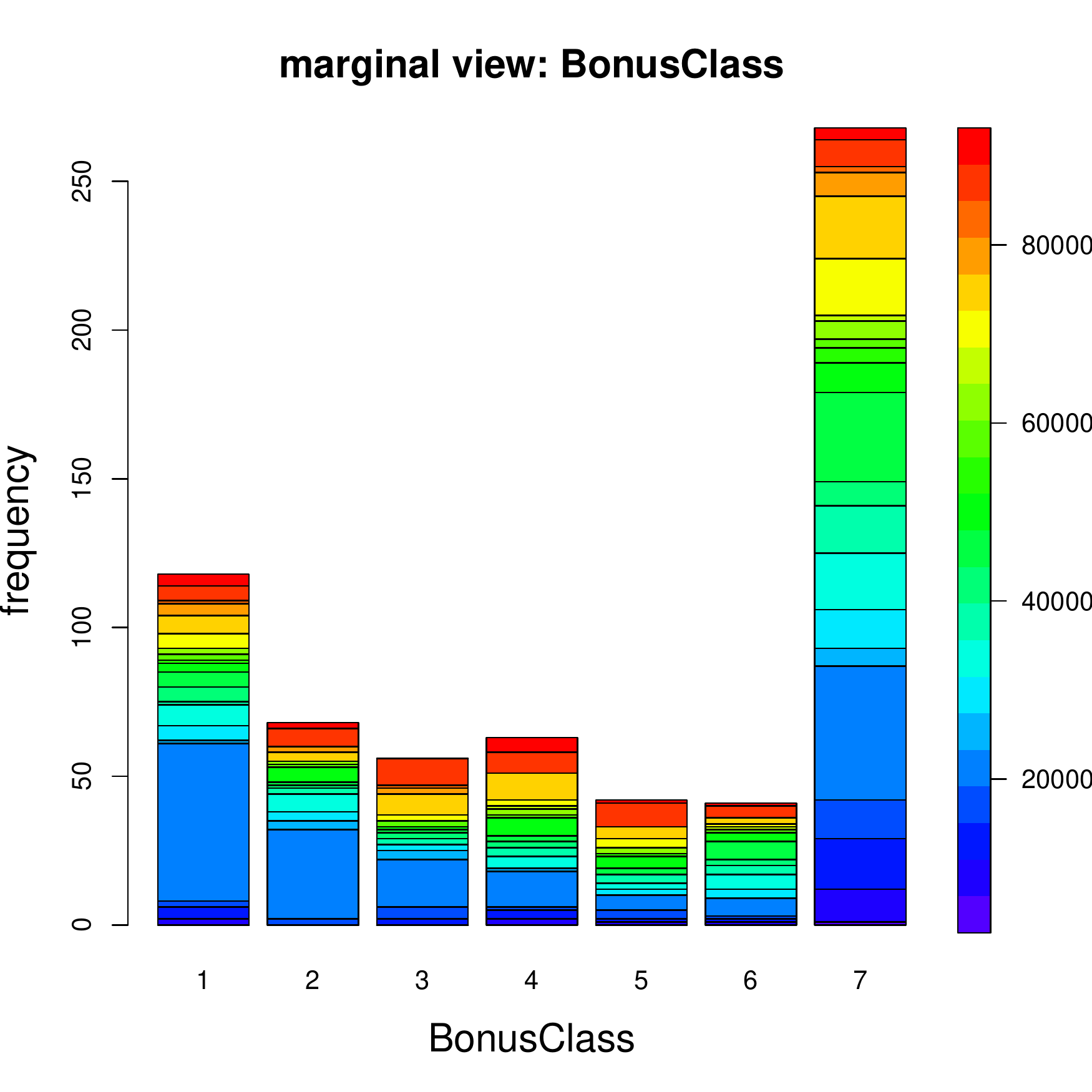}
\end{center}
\end{minipage}

\end{center}
\vspace{-.7cm}
\caption{Marginal view of the isotonically recalibrated gamma FFNN model of Table \ref{results gamma 1} of the 6 considered
  covariate components {\tt OwnerAge}, {\tt Gender}, {\tt Zone}, {\tt RiskClass}, {\tt VehAge}, {\tt BonusClass}.}
\label{Marginal plots Example 2}
\end{figure}

\begin{figure}[htb!]
\begin{center}
\begin{minipage}[t]{0.31\textwidth}
\begin{center}
\includegraphics[width=\textwidth]{./Plots/Bar_BonusClass.pdf}
\end{center}
\end{minipage}
\begin{minipage}[t]{0.02\textwidth}
\begin{center}
~
\end{center}
\end{minipage}
\begin{minipage}[t]{0.31\textwidth}
\begin{center}
\includegraphics[width=\textwidth]{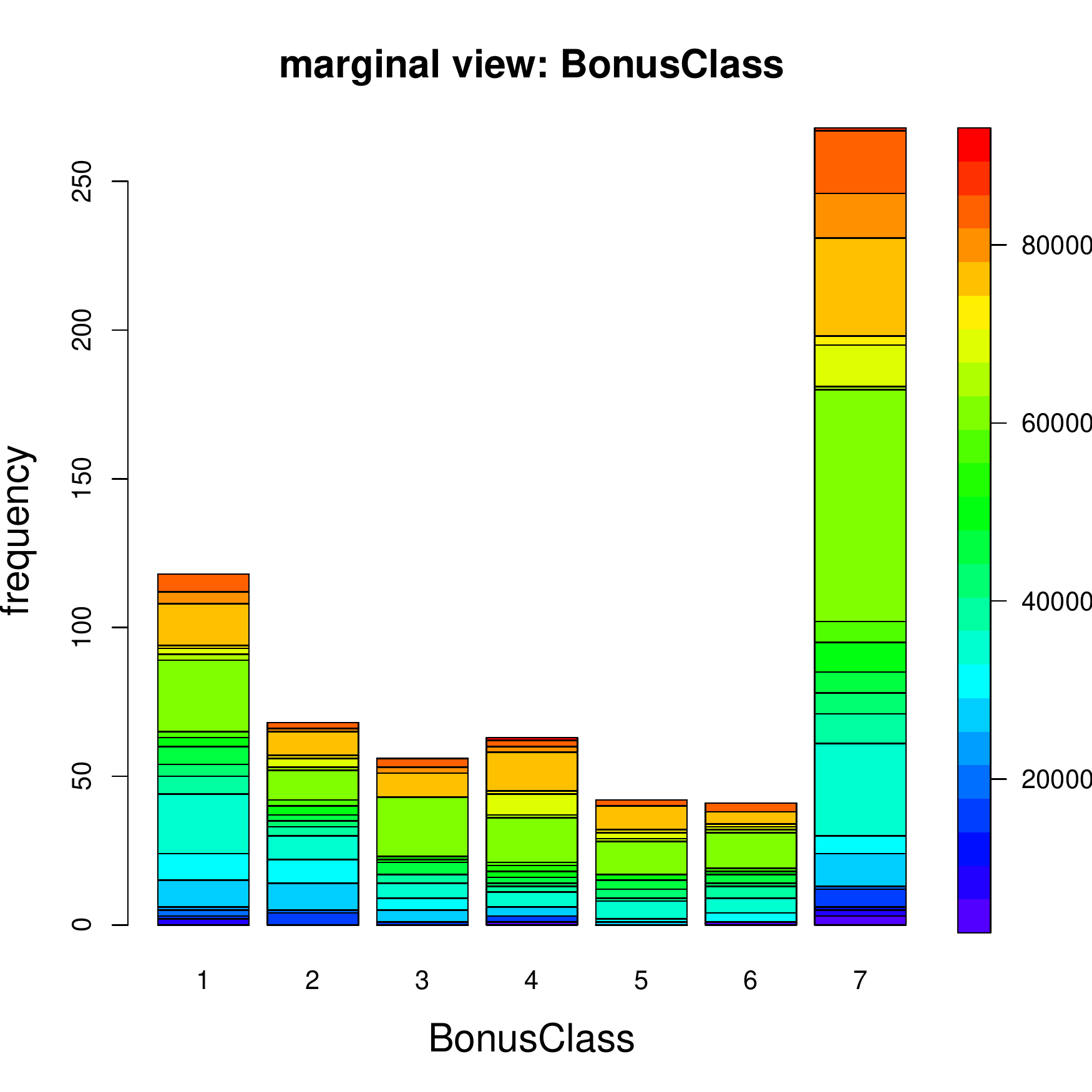}
\end{center}
\end{minipage}
\end{center}
\vspace{-.7cm}
\caption{Marginal view of the isotonically recalibrated gamma FFNN model (lhs) and
the isotonically recalibrated gamma GLM (rhs) for the covariate components {\tt BonusClass}.}
\label{Marginal plots Example 2B}
\end{figure}

In Figure \ref{Marginal plots Example 2}, we illustrate the resulting marginal plots if we project the estimated
values $\widehat{\bmu}$ of the isotonic recalibration to the corresponding covariate values, i.e., this is the
marginal view of the resulting covariate space partition \eqref{eq:def_partition}. For a low complexity number
$K((y_i,\widehat{\pi}(\bx_i))_{i=1}^n)$ this can be interpreted nicely. We see relevant differences in the distributions of the colors across the different covariate levels of {\tt OwnerAge}, {\tt Zone}, {\tt RiskClass} and {\tt VehAge}. This
indicates that these variables are important for explaining claim sizes, with the reservation that this marginal view ignores
potential interactions. For the variable {\tt Gender} we cannot make any conclusion as the gender balance
inequality is too large. The interpretation of {\tt BonusClass} is less obvious. In fact, from the gamma
GLM we know that {\tt BonusClass} is not significant, see \cite[Table 5.13]{WM2023}. This is because the {\tt BonusClass}
is related to collision claims, whereas our data studies comprehensive insurance that excludes collision claims. Figure \ref{Marginal plots Example 2B} shows the marginal view of the isotonically recalibrated gamma FFNN (lhs) and the gamma GLM (rhs) for the covariate {\tt BonusClass}. As mentioned,
{\tt BonusClass} is not significant in the gamma GLM, and it seems from the figure that, indeed, the color distribution across the different levels is rather similar for both models.

Clearly, the {\tt VehAge} is the most important variable showing the picture that claims on new motorcycles
are more expensive. There are substantial differences in claim size distributions between the zones, {\tt Zone 1} being the three largest cities
of Sweden having typically more big claims. {\tt RiskClass} corresponds to the size of the motorcycle which interacts with the {\tt OwnerAge}, the {\tt VehAge}
and the {\tt Zone}, and it is therefore more difficult to interpret as we have relevant interactions between these variables.

\begin{figure}[htb!]
\begin{center}
\begin{minipage}[t]{0.4\textwidth}
\begin{center}
\includegraphics[width=\textwidth]{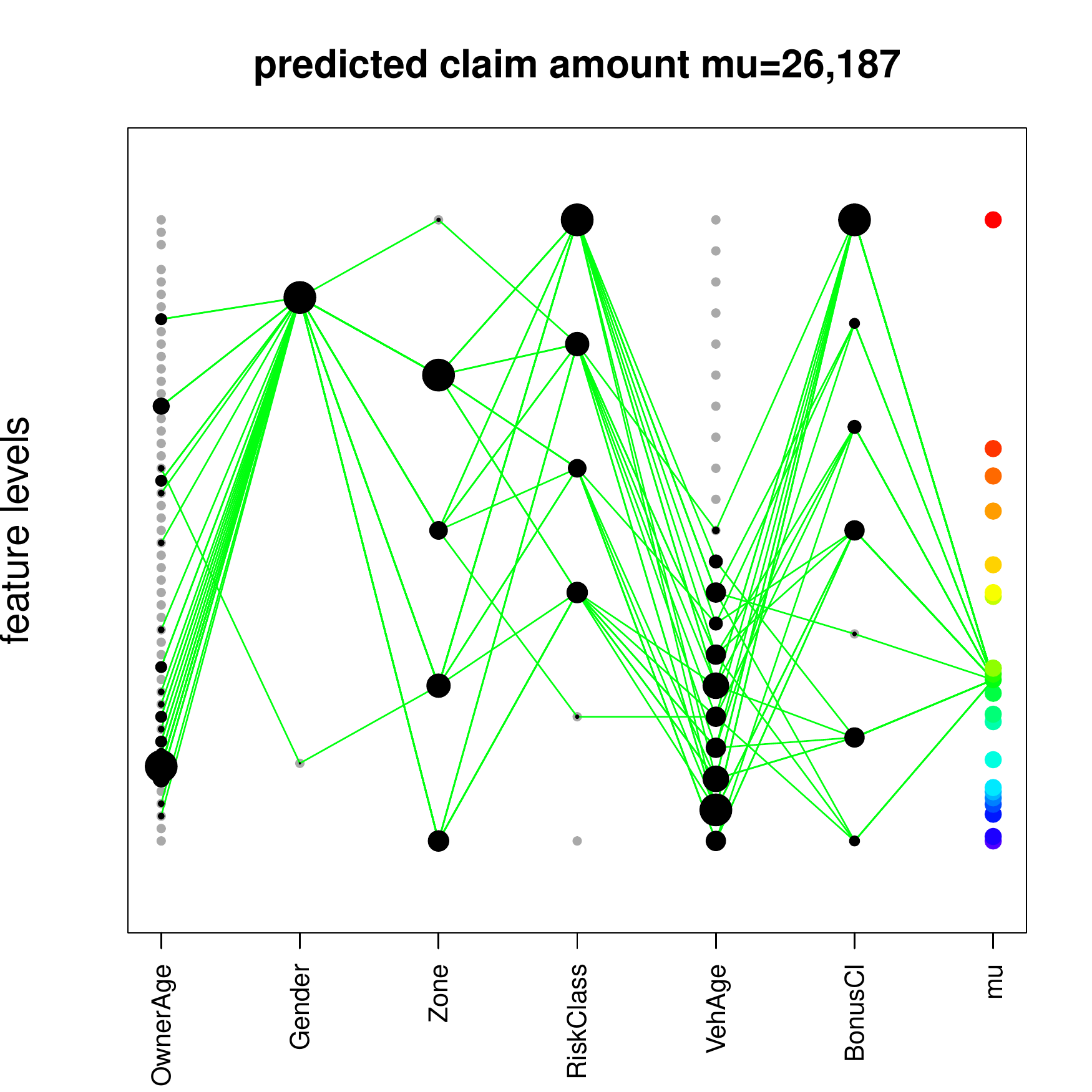}
\end{center}
\end{minipage}
\begin{minipage}[t]{0.4\textwidth}
\begin{center}
\includegraphics[width=\textwidth]{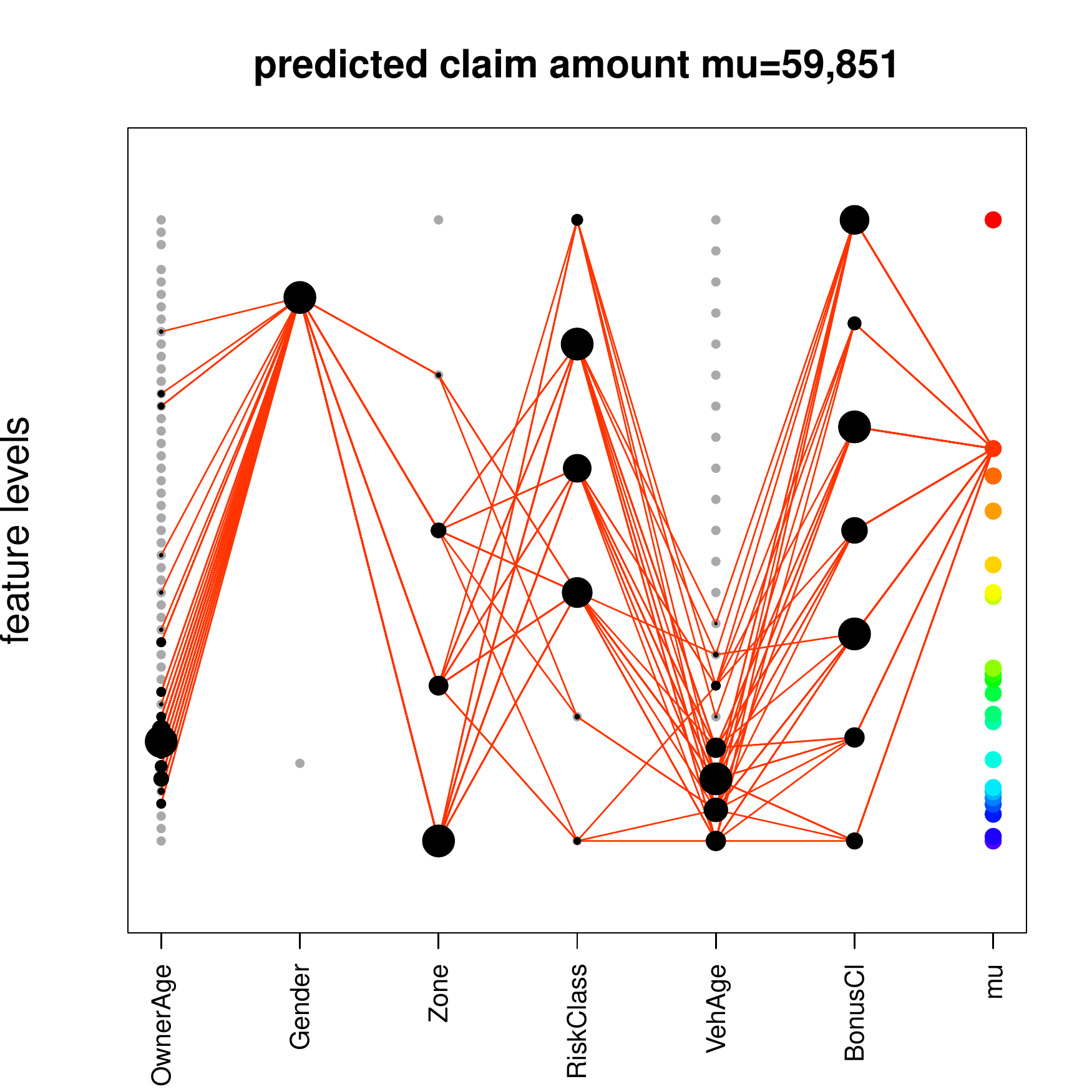}
\end{center}
\end{minipage}
\end{center}
\vspace{-.7cm}
\caption{Partition $({\cal X}_k)_{k=1,\dots,K}$ of the covariate space ${\cal X}$
w.r.t.~the isotonic recalibration for two selected values of $k=12, 21$.}
\label{Feature space partition}
\end{figure}

Figure \ref{Feature space partition} gives an illustration of the partition
$({\cal X}_k)_{k=1, \dots, K}$ of the 6-dimensional covariate space ${\cal X}$
w.r.t.~the isotonic recalibration $(\widehat{\mu}_{i_k})_{k = 1, \dots, K}$
for two selected values of $k$. The lines connect all the covariate components in $\bx$
that are observed within the data $(\bx_i)_{1\le i \le n}$ for a given value $\widehat{\mu}_{i_k}$, and the
size of the black dots illustrates how often a certain covariate level is observed.
E.g., the figure on the right-hand side belongs to the second largest
claim prediction $\widehat{\mu}_{i_{K-1}}=59,851$. For this expected response level,
the {\tt OwnerAge} is comparably small (around 25 years), everyone is {\tt Male} mostly living in 
{\tt Zone} 1 (three biggest cities of Sweden), having a motorcycle of a higher {\tt RiskClass} with a small {\tt VehAge}. Similar conclusions can be drawn for the other
parts ${\cal X}_k$ of the covariate space ${\cal X}$, thus, having a low complexity
number $K((y_i,\widehat{\pi}(\bx_i))_{i=1}^n)$ enables to explain the regression model.


\section{Conclusions}
\label{sec: Conclusions}
We have tackled two problems. First, we have enforced that the regression model fulfills the auto-calibration property by applying an isotonic recalibration to the ranks of a fitted (first) regression model. This isotonic recalibration does not involve any hyperparameters, but it solely assumes that the ranks from the first regression model are (approximately) correct. Isotonic regression has the property that the complexity of the resulting (non-parametric) regression function is small in low signal-to-noise ratio problems. Benefiting from this property, we have shown that this leads to explainable regression functions because a low complexity is equivalent to a coarse partition of the covariate space. In insurance pricing problems this is particularly useful, as we typically face a low signal-to-noise ratio in insurance claims data. We can then fit a complex (algorithmic) model to that data in a first step, and in a subsequent step we propose to auto-calibrate the first regression function using isotonic recalibration, which also leads to a substantial simplification of the regression function.



\bigskip

{\small 
\renewcommand{\baselinestretch}{.51}
}

\newpage

\appendix

\section{Appendix}
\subsection{Pool adjacent violators algorithm}
\label{appendix PAVA}
Minimization problem \eqref{isotonic regression} is a quadratic optimization problem with linear side constraints,
and it can be solved using the method of Karush--Kuhn--Tucker (KKT) \cite{KKT1, KKT2}.
We therefore consider the Lagrangian
\begin{equation*}
  L(\bmu, \boldeta) =  (\by - \bmu)^\top W (\by - \bmu) - \boldeta^\top A \bmu,
\end{equation*}
with Lagrange multiplier $\boldeta \in \R^{n-1}$. The KKT conditions are given by
\begin{eqnarray}
\label{KKT1} \b0&=& \nabla_{\bmu} L(\bmu, \boldeta) ~=~ -W(\by - \bmu) - A^\top \boldeta,
  \\\label{KKT2}
  \b0&\ge &\nabla_{\boldeta} L(\bmu, \boldeta) ~=~ -A \bmu,
  \\\label{KKT3}
  \b0& \le & \boldeta,\\\label{KKT4}
  \b0&=& \left(\eta_1(\mu_1-\mu_2), \ldots, \eta_{n-1}(\mu_{n-1}-\mu_n)\right)^\top.
\end{eqnarray}
The solution to these KKT conditions \eqref{KKT1}-\eqref{KKT4} provides the isotonic estimate $\widehat{\bmu}$.
This solution can be found by the PAV algorithm. The main idea is to compare raw
estimates $(\widetilde{\mu}_i)_i$. If we have an adjacent pair with $\widetilde{\mu}_i>\widetilde{\mu}_{i+1}$, it
violates the monotonicity constraint. Such pairs are recursively merged (pooled) to a block with an identical estimate, and
iterating this pooling of adjacent pairs and blocks, respectively, that violate the monotonicity constraint,
yields the PAV algorithm.

\hrulefill

{\sc Pool Adjacent Violators (PAV) Algorithm}

\vspace{-1ex}
\hrulefill
\begin{itemize}
\item[(0)] Initialize the algorithm $\widehat{\bmu}^{(0)}=\by$ and define the blocks
  ${\cal I}^{(0)}_k = \{k\}$ for $k=1,\ldots,  K^{(0)}=n$.
\item[(1)] Iterate for $t\ge 0$:
  \begin{itemize}
  \item[(a)] If $\widehat{\bmu}^{(t)}$ fulfills KKT condition \eqref{KKT2}
     go to item (2), otherwise go to the next step (1b).
  \item[(b)] Select an index $i=1,\ldots, n$ with $\widehat{\mu}_i^{(t)}>\widehat{\mu}_{i+1}^{(t)}$, merge the two adjacent
    blocks with $i \in {\cal I}_k^{(t)}$ and $i+1 \in {\cal I}_{k+1}^{(t)}$, and leave all other blocks
    unchanged. This
    provides the new blocks ${\cal I}_k^{(t+1)}$ with $k=1, \ldots,  K^{(t+1)}=K^{(t)}-1$.
    \item[(c)] Set on each block
$k=1,\dots, K^{(t+1)}$ and
    for all indices $i \in {\cal I}_k^{(t+1)}$ the
    new estimates 
    \begin{equation}\label{PAVA block estimate A}
      \widehat{\mu}_i^{(t+1)}= \frac{1}{\sum_{j \in {\cal I}_k^{(t+1)}}w_j}\,\sum_{j \in {\cal I}_k^{(t+1)}}w_jy_j.
      \end{equation}
    \item[(d)] Increase $t\mapsto t+1$.  
    \end{itemize}
  \item[(2)] Set the isotonic regression estimate $\widehat{\bmu}=\widehat{\bmu}^{(t)}$ and merge adjacent blocks
    ${\cal I}_k^{(t)}$ and ${\cal I}_{k+1}^{(t)}$ if we have the same estimates
    $\widehat{\mu}_i$ on these blocks. Return the resulting partition of ${\cal I}$ denoted by $({\cal I}_k)_{k=1,\dots,K}$
    and $\widehat{\bmu}$.
\end{itemize}
\vspace{-1ex}
\hrulefill

\begin{rems}[{\normalfont PAV algorithm interpretation}]\normalfont~
  \begin{itemize}
  \item[(0)] We initialize with the unconstraint optimal solution, and setting $\boldeta^{(0)}=0$ ensures that the
    KKT conditions \eqref{KKT1}, \eqref{KKT3} and \eqref{KKT4} are fulfilled, thus, only the monotonicity \eqref{KKT2} is not
    necessarily fulfilled.
  \item[(1a)] We identify a pair $\widehat{\mu}_i^{(t)}>\widehat{\mu}_{i+1}^{(t)}$ that violates the monotonicity
    constraint \eqref{KKT2}. This pair needs to belong to two adjacent blocks ${\cal I}_k^{(t)}$ and ${\cal I}_{k+1}^{(t)}$
    because within blocks we have constant estimates \eqref{PAVA block estimate}. We merge these two adjacent blocks
    to ${\cal I}_k^{(t+1)}={\cal I}_k^{(t)} \cup {\cal I}_{k+1}^{(t)}$, which reduces the number of blocks
    $K^{(t)}$ by 1.
  \item[(1b)] We set on each block the constant estimate \eqref{PAVA block estimate A} which satisfies the monotonicity
    constraint \eqref{KKT2} within blocks, and also \eqref{KKT4} is naturally fulfilled in this block.
    Conditions \eqref{KKT1} and \eqref{KKT3} are achieved by changing the Lagrange
    parameter $\boldeta^{(t)} \mapsto \boldeta^{(t+1)} \ge \b0$ correspondingly to account for the change in mean estimates
    \eqref{PAVA block estimate A} in \eqref{KKT1}.
  \item[(1c)] On a sample of size $n$, this algorithm can be iterated at most $n-1$ times, thus, the algorithm
    will terminate.
  \item[(2)] Since we have for $i \in {\cal I}_k^{(t)}$ and $i+1 \in {\cal I}_{k+1}^{(t)}$ the inequality
    $\widehat{\mu}_i^{(t)}\le \widehat{\mu}_{i+1}^{(t)}$, the last step is to ensure that the resulting blocks are
    maximal by merging blocks where we do not have a strict inequality in the corresponding estimates.
  \end{itemize}
\end{rems}  

\subsection{Proof of Theorem \ref{signal-to-noise theorem}}
{\Beweis
{\bf Proof of Theorem \ref{signal-to-noise theorem}.}
  For given responses $\by=\bY_\sigma(\omega)$, the solution to \eqref{isotonic regression} gives the partition \eqref{discrete interval}
  of the index set ${\cal I}$ with empirical weighted averages \eqref{PAVA block estimate} on the blocks ${\cal I}_k$.
  These empirical weighted averages satisfy $\widehat{\mu}_{i_k} < \widehat{\mu}_{i_{k+1}}$ for all $k=1, \ldots,  K(\bY)-1$,
  because the blocks ${\cal I}_k$ have been chosen maximal. We now consider how these blocks are constructed
  in the PAV algorithm. Suppose that we are in iteration $t\ge0$, and in this iteration of the PAV algorithm, we merge the two adjacent
  blocks ${\cal I}_k^{(t)}$ and ${\cal I}_{k+1}^{(t)}$ because $\widehat{\mu}_i^{(t)}>\widehat{\mu}_{i+1}^{(t)}$
  for $i \in {\cal I}_k^{(t)}$ and $i+1 \in {\cal I}_{k+1}^{(t)}$. We analyze this inequality
  \begin{equation*}
    \frac{1}{\sum_{j \in {\cal I}^{(t)}_k}w_j}\,\sum_{j \in {\cal I}^{(t)}_k}w_jy_j=    \widehat{\mu}_i^{(t)}
    ~>~\widehat{\mu}_{i+1}^{(t)}=\frac{1}{\sum_{l \in {\cal I}^{(t)}_{k+1}}w_l}\,\sum_{l \in {\cal I}^{(t)}_{k+1}}w_ly_l.
    \end{equation*}
We use the location-scale structure \eqref{PAVA block estimate} which gives us the equivalent condition
  \begin{equation*}
    \frac{1}{\sum_{j \in {\cal I}^{(t)}_k}w_j}\,\sum_{j \in {\cal I}^{(t)}_k}w_j\left(\mu_j+ \sigma \epsilon_j(\omega)\right)
    ~>~\frac{1}{\sum_{l \in {\cal I}^{(t)}_{k+1}}w_l}\,\sum_{l \in {\cal I}^{(t)}_{k+1}}w_l\left(\mu_l+ \sigma \epsilon_l(\omega)\right).
    \end{equation*}
    Since for any indices $j \in {\cal I}^{(t)}_k$ and $l \in {\cal I}^{(t)}_{k+1}$ we have $\mu_j \le \mu_l$, it follows
    that the previous condition for merging the two
    adjacent blocks ${\cal I}_k^{(t)}$ and ${\cal I}_{k+1}^{(t)}$ in iteration $t$ of the PAV algorithm reads as
    \begin{equation}\label{point-wise}
   \sigma \left[   
    \frac{\sum_{j \in {\cal I}^{(t)}_k}w_j\epsilon_j(\omega) }{\sum_{j \in {\cal I}^{(t)}_k}w_j}-
    \frac{\sum_{l \in {\cal I}^{(t)}_{k+1}}w_l \epsilon_l(\omega)}{\sum_{l \in {\cal I}^{(t)}_{k+1}}w_l}\right]
~> ~      \frac{\sum_{l \in {\cal I}^{(t)}_{k+1}}w_l \mu_l}{\sum_{l \in {\cal I}^{(t)}_{k+1}}w_l}-
 \frac{\sum_{j \in {\cal I}^{(t)}_k}w_j\mu_j}{\sum_{j \in {\cal I}^{(t)}_k}w_j}
~\ge~ 0.
    \end{equation}
    The important observation is that if this condition is fulfilled for scale parameter $\sigma>0$, then it will
    also be fulfilled for any bigger scale parameter $\sigma' > \sigma>0$ (pointwise
in $\omega \in \Omega$). Thus, any pooling that happens for $\sigma$
    also happens for $\sigma' > \sigma>0$. 
Since this is pointwise on the underlying probability space $(\Omega, {\cal F}, \p)$, it
shows that $\E[K(\bY)]$ is decreasing in $\sigma>0$.

Suppose now that the distribution of $\bepsilon$ has full support on $\R^n$. Then, the event $A_\sigma=\{K(\bY_\sigma)=n\}$
 occurs with positive probability, i.e.,
    \begin{eqnarray*}
      0 &<& \p[A_\sigma] ~=~ \p \left[ K(\bY_\sigma)=n \right]
      \\&=& \p \left[Y_1<Y_2< \ldots <Y_n \right]
      \\&=& \p \left[\mu_1+ \sigma \epsilon_1 < \mu_2+ \sigma \epsilon_2< \ldots <\mu_n + \sigma \epsilon_n \right].      
    \end{eqnarray*}

Consider
\begin{eqnarray*}
  A_\sigma &=& \left\{\mu_1+ \sigma \epsilon_1 < \mu_2+ \sigma \epsilon_2< \ldots <\mu_n + \sigma \epsilon_n
               \right\}
  \\&=& \bigcap_{k=2}^n \left\{\mu_{k-1}+ \sigma \epsilon_{k-1} < \mu_k+ \sigma \epsilon_k \right\}
  \\&=& \left\{\sigma \left(\epsilon_{1}-\epsilon_2\right) < \mu_2-\mu_{1} \right\}
         \cap \bigcap_{k=3}^n \left\{\mu_{k-1}+ \sigma \epsilon_{k-1} < \mu_k+ \sigma \epsilon_k \right\}.
  \end{eqnarray*}
We focus on the first event on the right-hand side. Note that $\mu_2 - \mu_1>0$, hence 
\[
\left\{\bepsilon \in \R^n :\epsilon_{1}-\epsilon_2 < \frac{\mu_2-\mu_{1}}{\sigma} \right\}
\]
describes an open half space in $\R^n$ containing the origin and with bounding hyperplane that moves further away from the origin when decreasing $\sigma$. Overall, the set $\tilde{A}_\sigma \subset \R^n$ of values of $\bepsilon$ in $A_\sigma$ is a non-empty open polyhedron containing the origin that scales with $\sigma$, that is, $\tilde{A}_\sigma = (1/\sigma) \tilde{A}_1$. Therefore, since the distribution of $\bepsilon$ has full support, the probability $\p[A_\sigma]$ is strictly decreasing in $\sigma$.


\EndProof}

\end{document}